\documentclass[]{aa}  

\usepackage{graphicx}
\usepackage{txfonts}

\usepackage[position=top]{subcaption}
\usepackage[export]{adjustbox}

\def\kms{\hbox{km$\;$s$^{-1}$}}
\def\ms{\hbox{m$\;$s$^{-1}$}}

\def\Halpha{\mbox{H\hspace{0.1ex}$\alpha$}}

\def\CaIR{\ion{Ca}{ii}~8542\,\AA}
\def\CaK{\ion{Ca}{ii}~K}
\def\CaH{\ion{Ca}{ii}~H}
\def\logtau{$\log\tau_{500}$}

\begin{document}

    \title{Umbral chromospheric fine structure and umbral flashes modelled as one: the corrugated umbra}

        \author{Vasco M.J. Henriques\inst{1}$^,$\inst{2},
                  %\and
          Chris J. Nelson\inst{3},
          %\and
          Luc H.M. Rouppe van der Voort\inst{1}$^,$\inst{2},
          \and 
           Mihalis Mathioudakis\inst{3}
          }

   \institute{Institute of Theoretical Astrophysics, University of Oslo, 
   P.O. Box 1029 Blindern, N-0315 Oslo, Norway
         \and
             Rosseland Centre for Solar Physics, University of Oslo, P.O. Box 1029 Blindern, N-0315 Oslo, Norway
                             \and
             Astrophysics Research Centre (ARC), School of Mathematics and Physics, Queen's University Belfast, BT7 1NN, Belfast, Northern Ireland, UK\\
             \email{vh@astro.uio.no}
             }

   \date{Received; accepted}
   
  \abstract

  \abstract
   {The chromosphere of the umbra of sunspots features an assortment of dynamic fine structures that are poorly understood and often studied separately. Small-scale umbral brightenings (SSUBs), umbral microjets, spikes or short dynamic fibrils (SDFs), and umbral dark fibrils are found in any observation of the chromosphere with sufficient spatial resolution performed at the correct umbral flash stage and passband. Understanding these features means understanding the dynamics of the umbral chromosphere.}
  % aims heading 
   {We aim to fully understand the dynamics of umbral chromosphere through analysis of the relationships between distinct observed fine features and aim to produce complete models that explain both spectral profiles and the temporal evolution of the features. We seek to relate such understanding to umbral flashes.}
  % methods heading 
   {We study the spatial and spectral co-evolution of SDFs, SSUBs, and umbral flashes in \CaIR{} spectral profiles. We produce models that generate the spectral profiles for all classes of features using non-local thermodynamic equilibrium (non-LTE) radiative transfer with a recent version of the NICOLE inversion code. }
  % results heading 
   {We find that both bright (SSUBs) and dark (SDFs) structures are described with a continuous feature in the parameter space that is distinct from the surroundings even in pixel-by-pixel inversions. We find a phase difference between such features and umbral flashes in both inverted line-of-sight velocities and timing of the brightenings. For umbral flashes themselves we resolve, for the first time in inversion-based semi-empirical modelling, the pre-flash downflows, post-flash upflows, and the counter-flows present during the umbral flash phase. We further present a simple time-dependent cartoon model that explains the dynamics and spectral profiles of both fine structure, dark and bright, and umbral flashes in umbral chromospheres.} 
   %conclusions
   {The similarity of the profiles between the brightenings and umbral flashes, the pattern of velocities obtained from the inversions, and the phase relationships between the structures all lead us to put forward that all dynamic umbral chromospheric structures observed to this date are a locally delayed or locally early portion of the oscillatory flow pattern that generates flashes, secondary to the steepening large-scale acoustic waves at its source. Essentially, SSUBs are part of the same shock or merely compression front responsible for the spatially larger umbral flash phenomenon, but out of phase with the broader oscillation.}

\titlerunning{SSUBs and short dynamic fibrils}
\authorrunning{Henriques et al.}
   \maketitle

\section{Introduction}

Sunspots manifest as dark regions, with spatial scales of tens of arcseconds and lifetimes of the order days. The umbrae of sunspots in the photosphere are often highly structured containing features such as umbral dots, dark-cored filaments, and lightbridges (see, for example, \citealt{2008ApJ...672..684R}) which are relatively stable over timescales of the order of minutes. Observations of sunspot umbrae in strong chromospheric spectral lines, however, reveal a different picture with umbral flashes, a remarkably dynamic large-scale oscillatory pattern with a periodicity of about 3~min \citep{1969SoPh....7..351B,1969SoPh....7..366W,2003A&A...403..277R} dominating time-series of images. The origin of these oscillations likely stems from the p-mode sub-photospheric oscillations which generate acoustic waves just above the acoustic cutoff frequency in the lower chromosphere and that are expected, from simulations, to steepen into shocks as they move higher through the chromosphere \citep[e.g.,][]{1984ApJ...277..874L,1984A&A...135..188T,1992ASIC..375..261L,2007ApJ...671.1005B,2010ApJ...722..888B,2010ApJ...722..131F,Felipe_2014}. 
In recent years much progress has been achieved on the large-scale understanding of such oscillations, and how they interact with the broad magnetic field structure of the sunspot. A large portion of this progress has been achieved through techniques such as power spectrum analysis (\citealt{2015ApJ...812L..15K, 2017ApJ...842...59J,2018A&A...617A..39F,2018ApJ...854..127C,2019A&A...627A.169F,2019ApJ...879...67C,2020NatAs...4..220J}) and we would refer the reader to the recent reviews by \cite{2015SSRv..190..103J} and \cite{2015LRSP...12....6K} for an overview of such line of work. Less understood are the local conditions during the shocks themselves, as well as the fine structure within umbral flashes, both horizontal fine structure and the vertical stratification at scales smaller than the wavelength of the acoustic waves at their source, especially when it comes to complete semi-empirical atmospheric models of the umbra. 

Fine structuring within umbral flashes was first suspected when \cite{2005ApJ...635..670C} observed abnormal Stokes V profiles in \ion{He}{i} $10830$ \AA\ spectra during umbral flashes. Direct observations of such fine structure were achieved soon after by \cite{2009ApJ...696.1683S} who found that dark slender absorption features were often present co-spatial to bright umbral flashes in narrow-band imaging sampling the \ion{Ca}{ii}~H line. This work was later confirmed and expanded on by \cite{2013A&A...557A...5H} who used filtergrams sampled in the same spectral region, and \cite{2015A&A...574A.131H} who confirmed that these events could be repeated over multiple flashes and that two classes of sizes likely existed, with some features having a match in \Halpha\ absorption structures. The smaller class of these structures were likely short dynamic fibrils which were first observed and characterised by \cite{2013ApJ...776...56R} using both the \ion{Ca}{ii}~8542 \AA\ and \Halpha{} spectral lines. Short dynamic fibrils (from now on SDFs) exhibit parabolic profiles in time-distance plots, with acceleration and deceleration at values that can significantly depart from solar gravity, a property consistent with their magneto-acoustic nature, as found for dynamic fibrils in general \citep{2006ApJ...647L..73H,2007ApJ...655..624D}. SDFs feature Doppler signals matching their ascent phase and their descent phase (i.e., blue-shifted followed by red-shifted spectral profiles). SDFs have also been described as spikes in \Halpha\ observations \citep{2014ApJ...787...58Y}, and can have remarkable extents (up to $1$~Mm) in height. 

In addition to absorption features, fine-scale chromospheric structuring can also be detected as localised brightenings, with sub-arcsecond lengths and sub-minute lifetimes, in sunspot umbrae. Such events were first reported by \cite{2013A&A...552L...1B} who interpreted them as microjets potentially driven by magnetic reconnection around umbral dots. \cite{2017A&A...605A..14N} expanded on this work finding that such small-scale umbral brightenings (hereafter referred to as SSUBs) were also visible in the \ion{Ca}{ii}~8542 \AA\ line and found that they did not appear to have the expected properties of jets (e.g., no parabolic evolution in lengths). Instead those authors found that these localised brightenings exhibited flash-like profiles in \ion{Ca}{ii}~8542 \AA\ spectra and occurred at the foot-points of SDFs during the latter's downflowing stage. Thus, it was deemed highly likely that SSUBs are not formed by an out-bursting jet but rather by localised compression shocks. It was proposed that either increased emission via a shock caused by the returning downflowing material impacting the at-rest umbral "floor" or a perturbation to a passing umbral flash front due to a denser downflowing atmosphere could explain the spectral, spatial, and temporal properties of SSUBs. More recently, \cite{2020MNRAS.493.3036B} reported that multiple types of localised brightenings could be detected in the chromosphere above an evolving pore, with some being potentially similar to SSUBs and some appearing to form due to magnetic reconnection above a lightbridge.

Large-scale waves generate transient flow fields that impact the formation of the spectral profiles of chromospheric lines. In addition to the flow-field generated by waves, external sources of flows can be locally present \citep{2015A&A...582A.116S,2016A&A...587A..20C,2018ApJ...859..158S,2020A&A...636A..35N}. Observations will be sensitive to all such processes. Semi-empirical modelling using non local thermodynamic equilibrium (non-LTE) inversions has found that upflowing solutions best model the moment of the umbral flash \citep{2000Sci...288.1398S,2000ApJ...544.1141S,2013A&A...556A.115D,2018A&A...619A..63J} and so has forward synthesis based on simulations \citep{Felipe_2014} where upflows are naturally generated as part of the steepening up-ward propagating wave. However, recent semi-empirical results have shown that umbral flash profiles can be reproduced with atmospheres dominated by strong downflows \citep{2017ApJ...845..102H,2019A&A...627A..46B,2020ApJ...892...49H}. All of these semi-empirical studies analysed the \CaIR{} spectral line using different versions of the Non-LTE Inversion COde using the Lorien Engine (\verb|NICOLE|; \citealt{2015A&A...577A...7S}), with \cite{2019A&A...627A..46B} also using the RH code \citep{Uitenbroek2001,Tiago2015RH} to synthesise \ion{Mg}{ii} UV lines, finding a good agreement between UV observations and their models. 

As the acoustic waves move into lower density media, increasing in amplitude, the presence of shocks are expected. Such shocks are visible as strong discontinuities in the snapshot models of the similarly formed \ion{Ca}{ii}~H grains \citep{1997ApJ...481..500C} and are also predicted from the simulations of acoustic waves steepening in the umbra by \citet{2010ApJ...722..888B} or \citet{Felipe_2014}. The presence of shocks is expected in semi-empirical modelling but such a discontinuous front has so far not been possible to be directly resolved. Using Doppler and line-width fitting in \ion{He}{i} $10830$ \AA\ observations, \cite{2019ApJ...882..161A} build a shock model using two homogeneous slabs (i.e. of constant properties throughout) that are consistent at the interface via the Rankine–Hugoniot relations to constrain density, velocity and pressure on both sides of the shock-front and estimate the heating deposited by such shock. Similarly \cite{2020ApJ...892...49H} use inverted atmospheres from \CaIR{} to test different shock models where the Rankine–Hugoniot relations are also used and find evidence for shocks at multiple heights in the atmosphere.  The downflow-dominated models for umbral flashes of \cite{2019A&A...627A..46B} were used as input to the shock-heating analysis of \cite{2019ApJ...882..161A} while \cite{2020ApJ...892...49H} rely on bulk upflowing or downflowing models. Resolving the flow structure during umbral flashes is thus becoming increasingly important to properly constrain their energy deposition via dissipation into heat within shocks at whatever heights and locations they are found to occur. Due to their localised nature in height, up-wardly propagating in time, umbral flashes have the potential to be used as probes of the local plasma conditions over a large range of heights, with notable results including the evidence for local adiabatic expansion and a corresponding magnetic field reduction upon their passage \citep{2017ApJ...845..102H,2018ApJ...860...28H}. It is thus essential that further work is undertaken to understand the local hydrodynamical conditions during a flash and their line-formation.

In this paper, we investigate the properties of SSUBs using non-LTE inversions performed with \verb|NICOLE|. Additionally, we study the relationship between SSUBs and SDFs using their spectral co-variation in time. As both SDFs and SSUBs were found to occur across the entire umbra of the observed sunspots whenever they were studied \citep{2013ApJ...776...56R,2017A&A...605A..14N,2014ApJ...787...58Y} this work will open the window into the umbra of sunspots and its complex dynamics. 

\begin{figure*}[!ht]
\centering
\includegraphics[clip=true,trim=0.9cm 0.0cm 0.0cm 0.0cm]{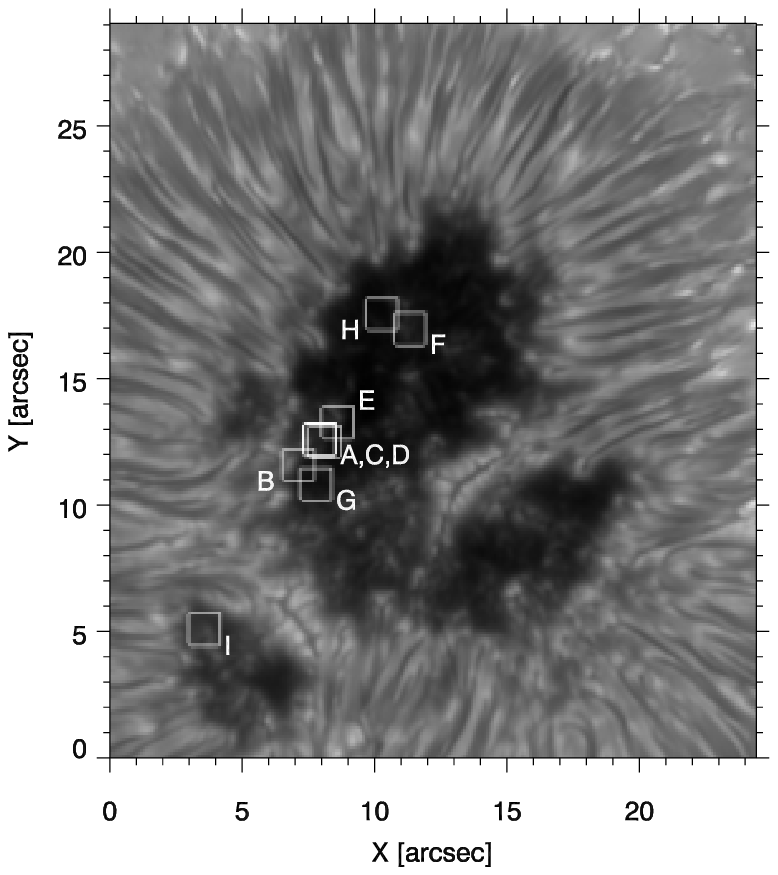}
\includegraphics[width=8.8cm,clip=true,trim=2cm 0.0cm 0.0cm 0.5cm]{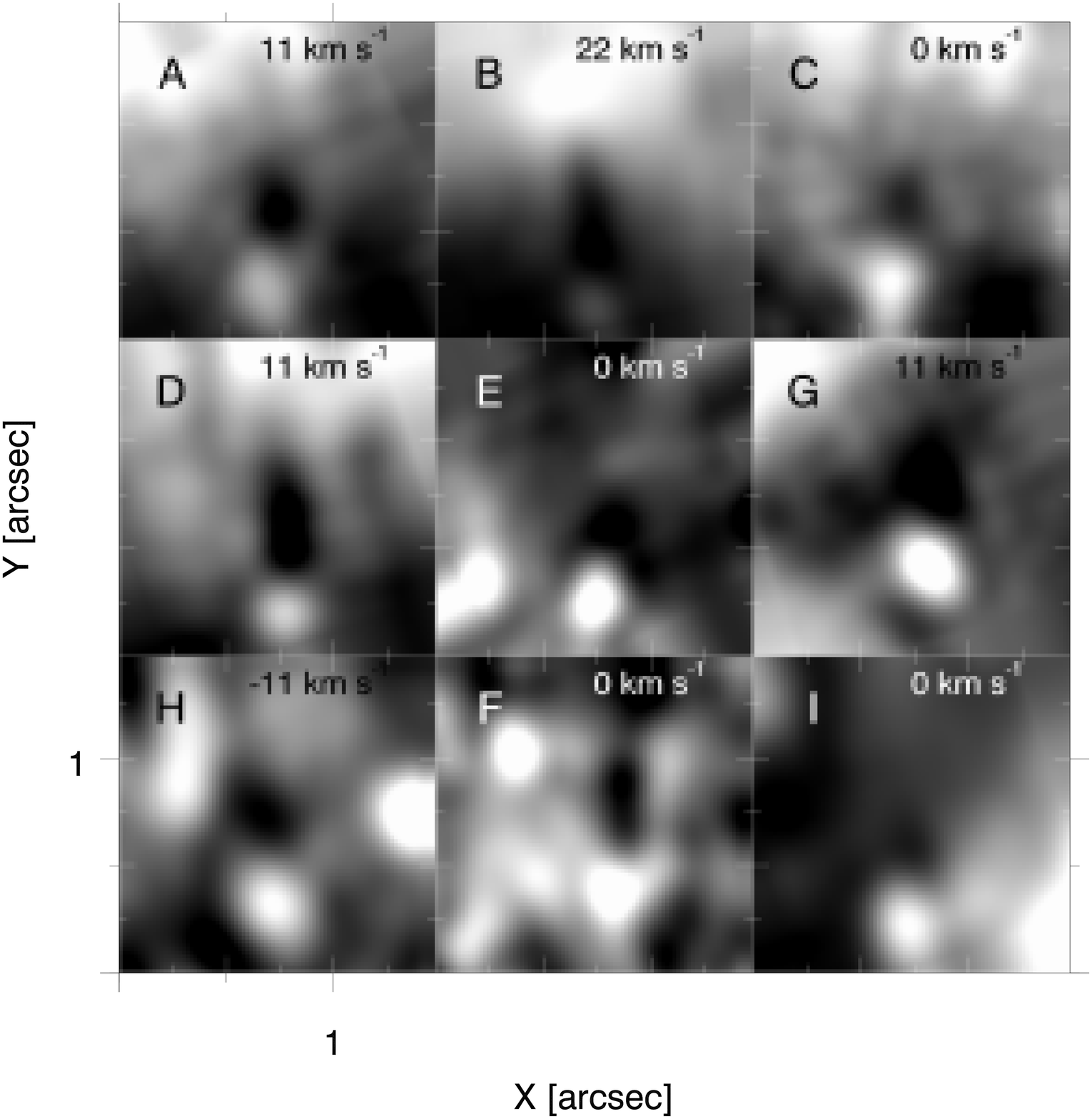}
\includegraphics[clip=true,trim=0.9cm 0.0cm 0.0cm 0.0cm]{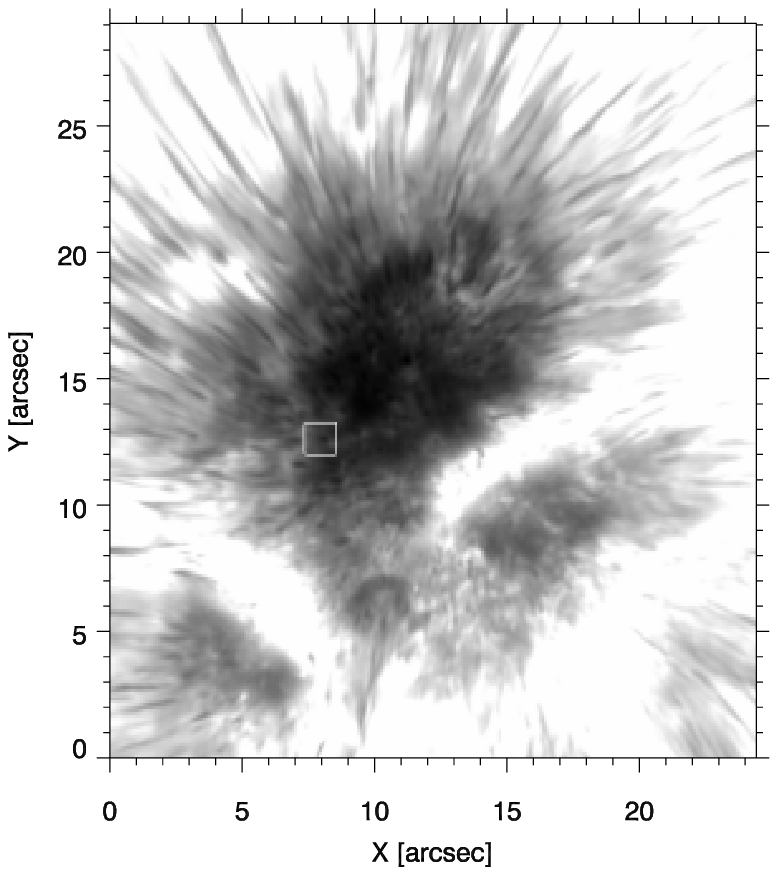}
\includegraphics[clip=true,trim=0.4cm 0.0cm 0.5cm 0.0cm]{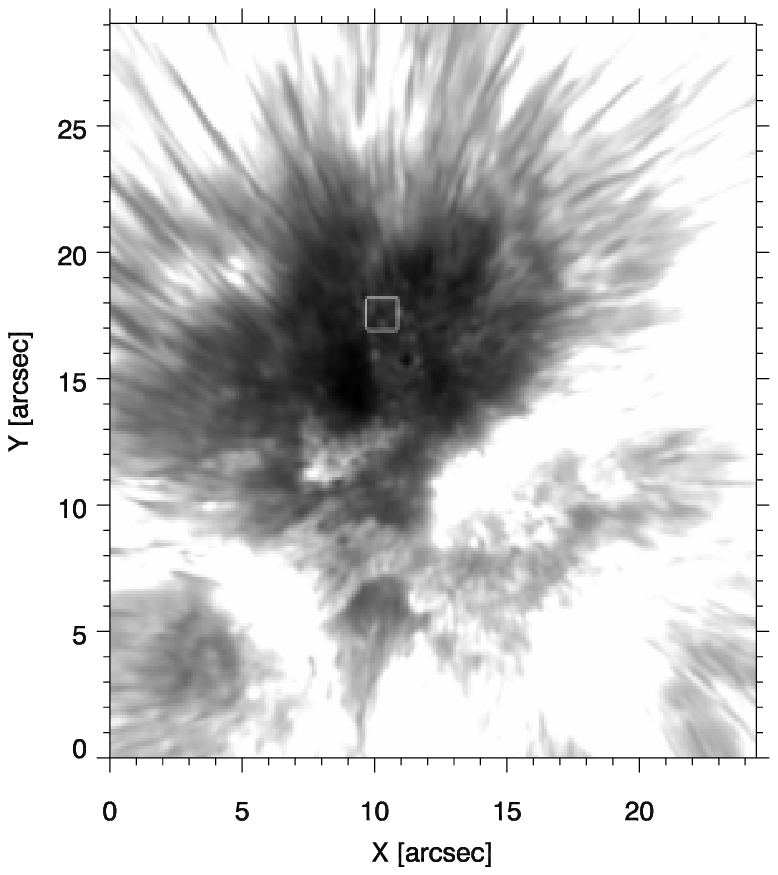}
\caption{Overview of the nine selected SDF/SSUB events. The top-left panel shows a context \ion{Ca}{ii}~8542 \AA\ far wing image ($-$942~m\AA) with the location of the events marked with small squares, delineating the exact areas that are analysed using inversions. The nine sub-images at the top-right show the SDF--SSUB pairs at their clearest visibility and are selected from Figs.~\ref{trimedpanels}, \ref{trimedpanels2} and \ref{trimedpanels3}. These correspond to the fields-of-view shown in the top-left panel with a rotation applied such that the SDF--SSUB pairs are aligned along the $y$-axis. In all cases the dark SDFs are seen just above the bright SSUBs. The central wavelength of each passband is written in units of Doppler shift from the core. The bottom panels show two SDF--SSUB pairs in context. Bottom left is an image of the line-core, scaled such that 30\% of the pixels are saturated. The box highlights SDF--SSUB pair A. The bottom right panel is an image taken at $+$145~m\AA\ from line-core, with the same 30\% saturation, where SDF--SSUB pair H is highlighted with a box. This latter selection perfectly matches the respective image in the top right panel as no rotation was applied.} 
\label{context}
\end{figure*}

\section{Data acquisition and analysis}

\subsection{Observations}

The observations analysed here are the same as those of \cite{2017ApJ...845..100R}, \cite{2017A&A...605A..14N} and \cite{2017ApJ...845..102H}. The main target of the observations and of the inversions was the largest umbra of the main sunspot in the active region NOAA 12121 which was sampled between 10:43 and 11:23 UT on the 28th of July 2014, when it was close to disk centre. The \ion{Ca}{ii}~H data used to identify the SSUBs were taken with the blue tower setup using its \ion{Ca}{ii}~H~1.1\AA\ FWHM interference filter centred at the core of the line. The  cadence was 1.4~s, the pixel scale 0\farcs034 and the resolution was close to the diffraction limit of 0\farcs1 at that wavelength. Further details on those observations, their reduction, and the SSUB detection methods are described in \cite{2017A&A...605A..14N}. We use a sub-set of eight SSUB events discussed in \cite{2017A&A...605A..14N}, who implemented a detection algorithm similar to that of \cite{2013A&A...552L...1B} in the \CaH\ channel, and one new event identified in the \CaIR{} line directly. Nine features corresponding to nine sub-fields were selected for analysis here. The selection of these nine events was conducted such that the SSUBs should be sufficiently easy to recognise visually in the \ion{Ca}{ii} $8542$ \AA\ spectra without the use of running differences.

The \ion{Ca}{ii}~8542 \AA\ data were acquired with the CRisp Imaging SpectroPolarimeter \citep[CRISP,][]{2006A&A...447.1111S,2008ApJ...689L..69S} instrument, at the Swedish 1-m Solar Telescope \citep[SST;][]{2003SPIE.4853..341S}. 
We re-processed the spectral scans as compared to the data analysed in \cite{2017A&A...605A..14N} and \cite{2017ApJ...845..102H} with the aim of producing a time series with an increased time resolution of 14~s per scan (previously 28~s per scan followed by 1~s spent in H-alpha core). Compared to those previous works the increased time resolution was achieved at the expense of polarimetric signal (due to the halving of the total integration time), but polarimetry was deemed to be a lower priority than the capture of fast evolving flow structures for this work. 
Adaptive optics were used, including an 85-electrode deformable mirror which is an upgrade of the system described in \cite{2003SPIE.4853..370S}. All data were reconstructed with Multi-Object Multi-Frame Blind Deconvolution \citep[MOMFBD;][]{2002SPIE.4792..146L,2005SoPh..228..191V}, using 80~Karhunen-Lo\`{e}ve modes sorted by order of atmospheric significance and $88\times88$ pixel subfields (only 52 modes were used in the previous papers from these data). 
A version of the completed reduction pipeline published by \cite{2015A&A...573A..40D} was used before and after MOMFBD. This includes the method described by \cite{2012A&A...548A.114H} for consistent alignment between the different liquid crystal (LC) states and wavelengths, with destretching performed as in \cite{1994ApJ...430..413S}. The pipeline includes compensation for the prefilter transmission profile. Post-pipeline the observations were normalised to the intensity of the continuum levels by fitting the "Fourier Transform Atlas" \ion{Ca}{ii}~8542~\AA\ profile \citep{1999SoPh..184..421N}, convolved with the theoretical double-cavity transmission profile of CRISP, to an average of the quiet-Sun profile computed from multiple scans as in \cite{2013A&A...556A.115D}.

\begin{figure*}[]
\begin{center}
\includegraphics{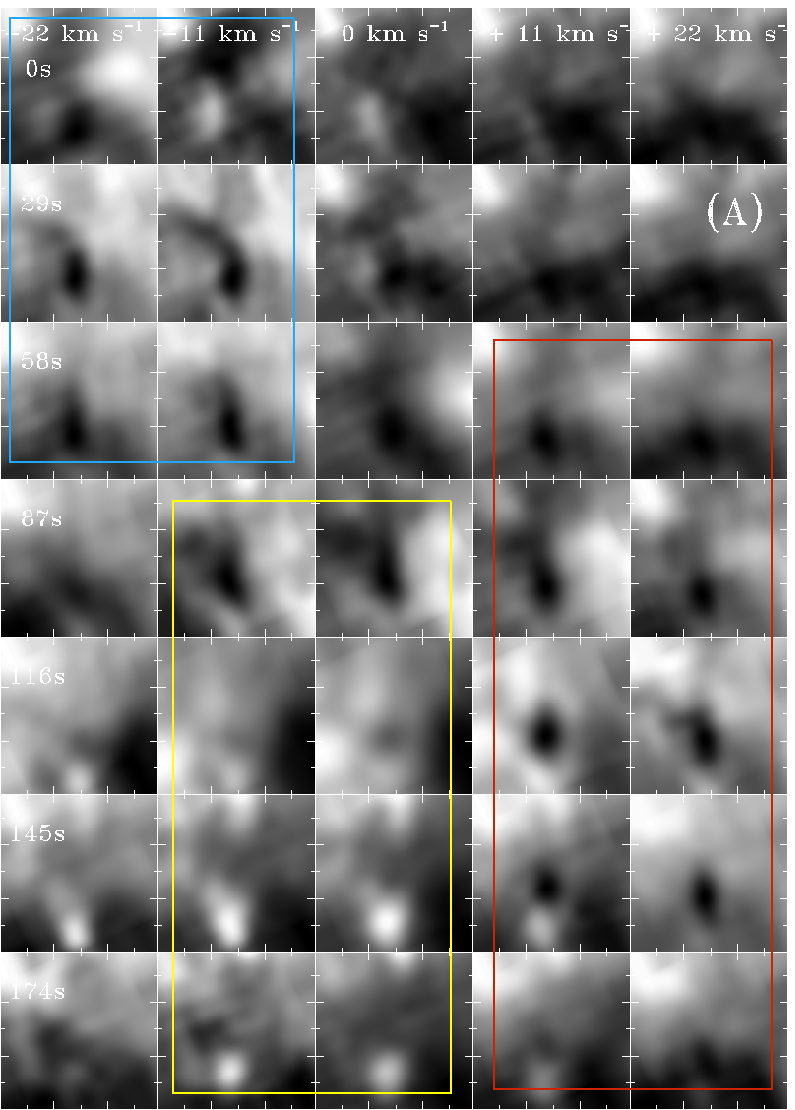} % A 
\hspace{0.5mm}
\includegraphics{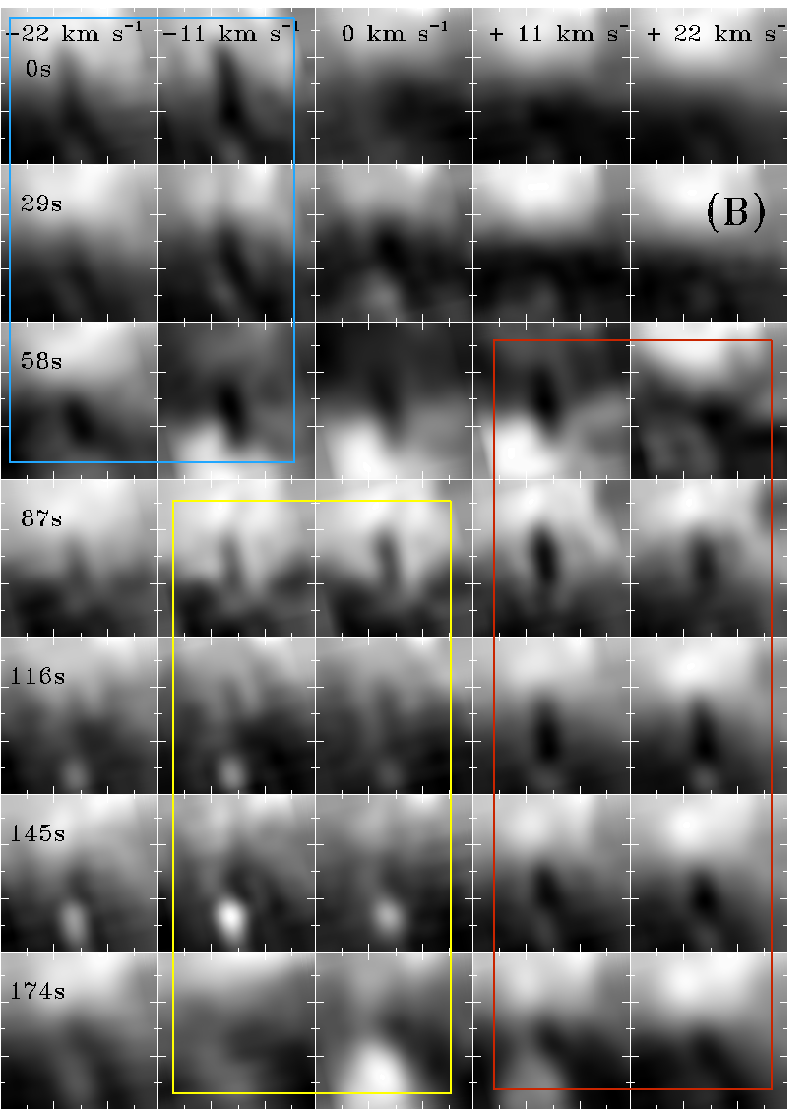} % B 
\vspace{2mm}
\includegraphics{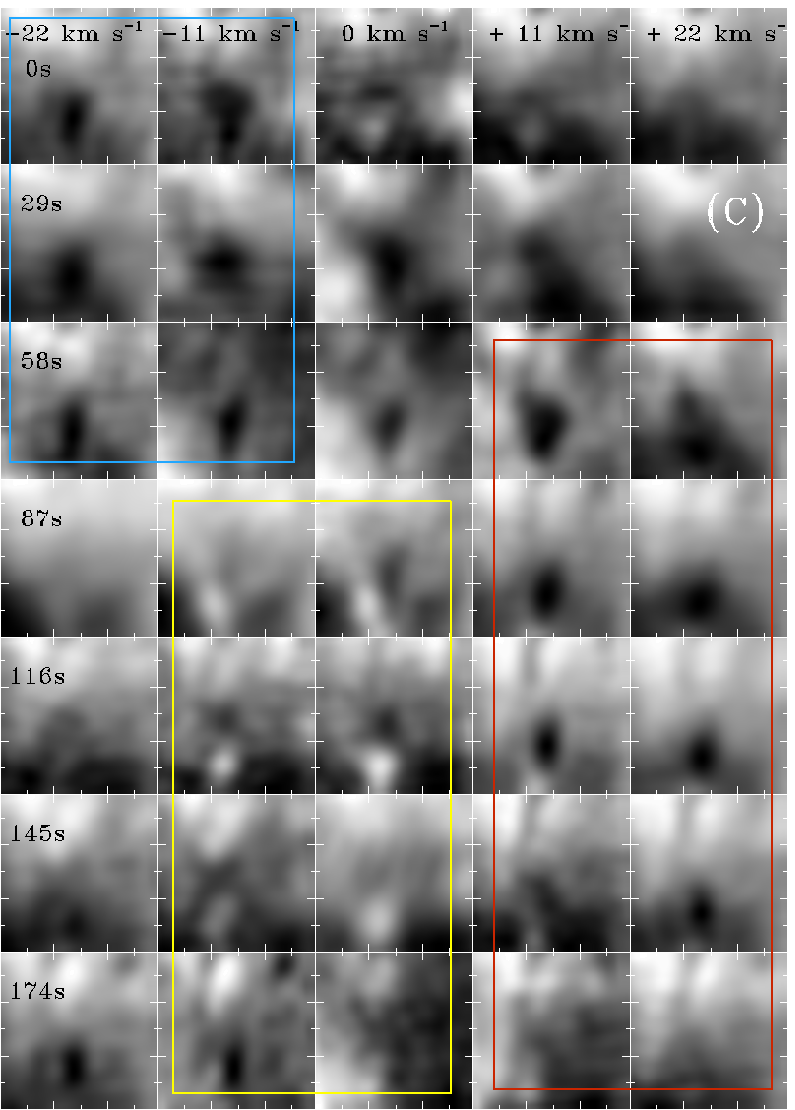} % C
\hspace{0.5mm}
\includegraphics{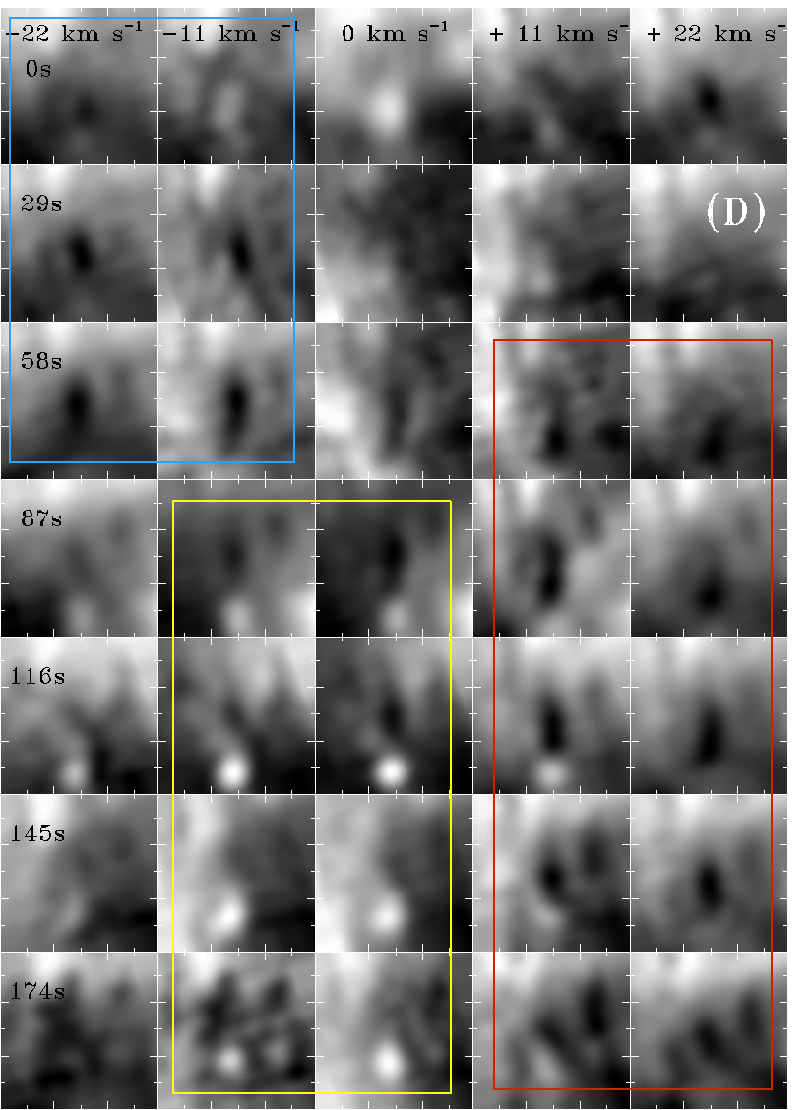} % D 
\end{center}
\caption{\footnotesize Evolution of four different SSUB/SDF pairs (events A to D). Each subfield is an image at a given bandpass, labelled in velocity units from line-core, and a given time in seconds from the approximate onset of the SDF. Thus left to right is blue to red-shift and top down is time evolution. The coloured boxes highlight different stages of evolution of the SSUB/SDF pairs as described in the text. The blue-box highlights the upflowing/growth stage of the SDF, red the downflowing stage, and yellow the corresponding SSUB flash. Major tickmarks every 0\farcs5.} 
\label{trimedpanels}
\end{figure*}

\begin{figure*}[]
\begin{center}
\includegraphics{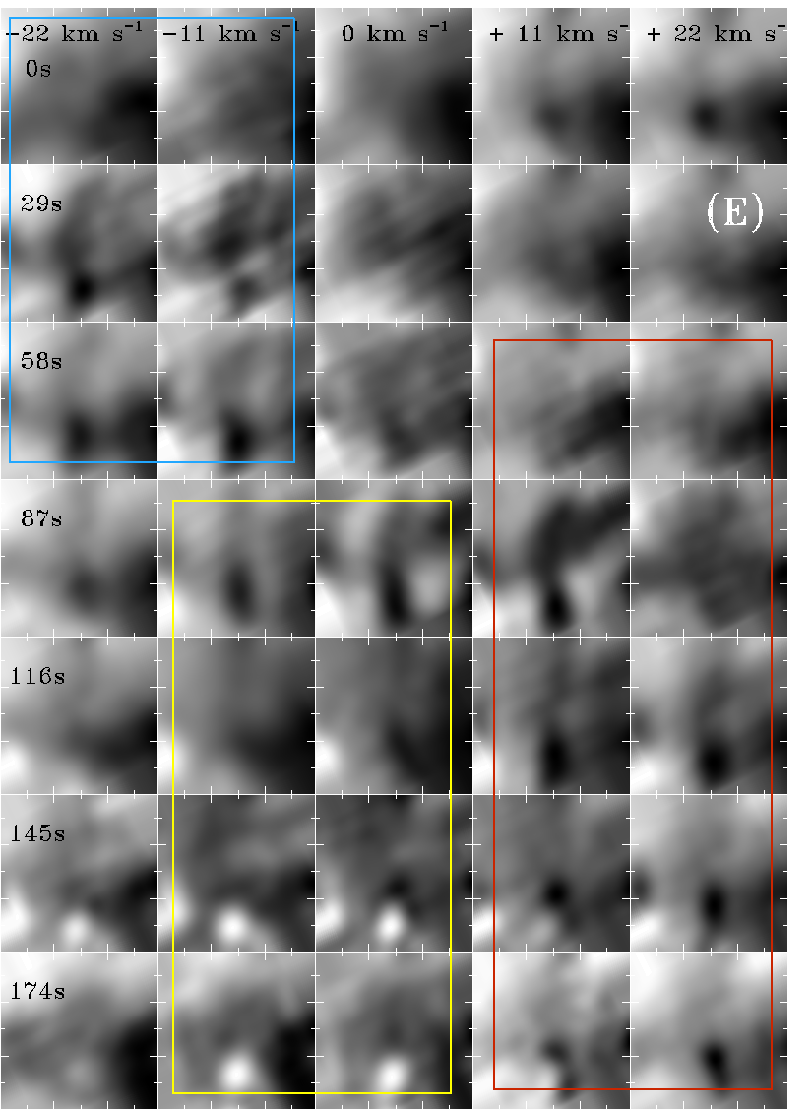} % E 
\hspace{0.5mm}
\includegraphics{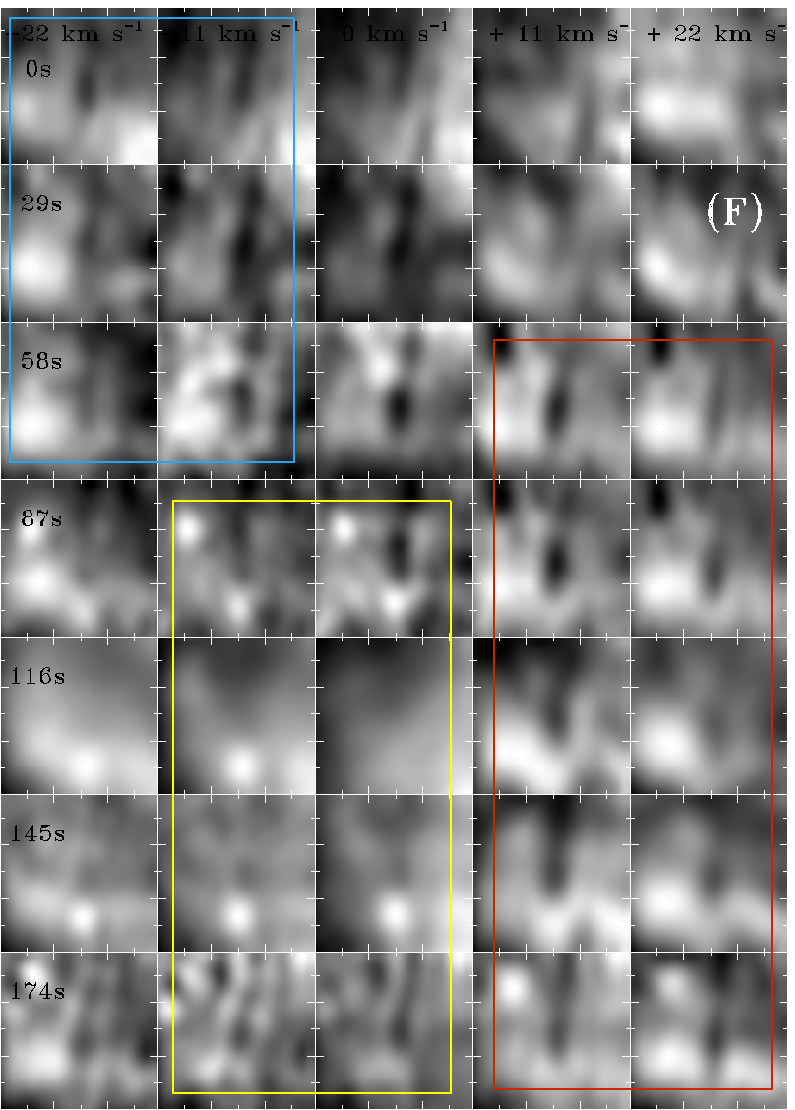} % F
\vspace{2mm}
\includegraphics{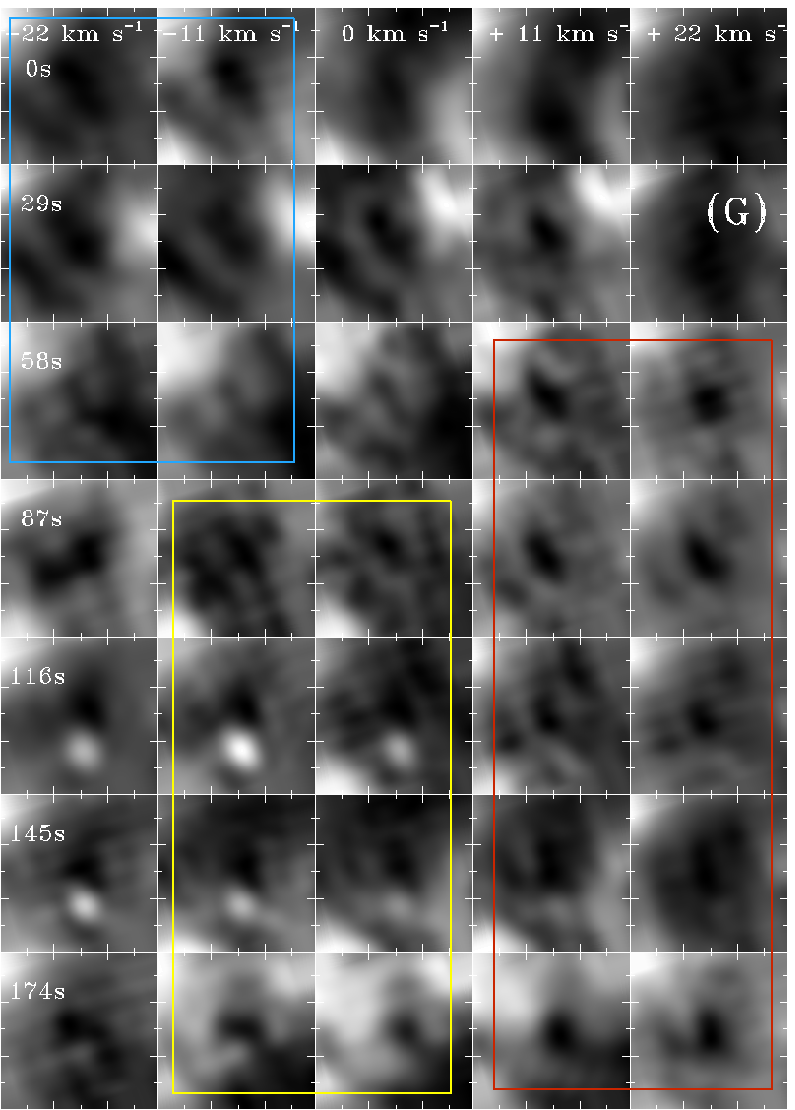} 
\hspace{0.5mm}
\includegraphics{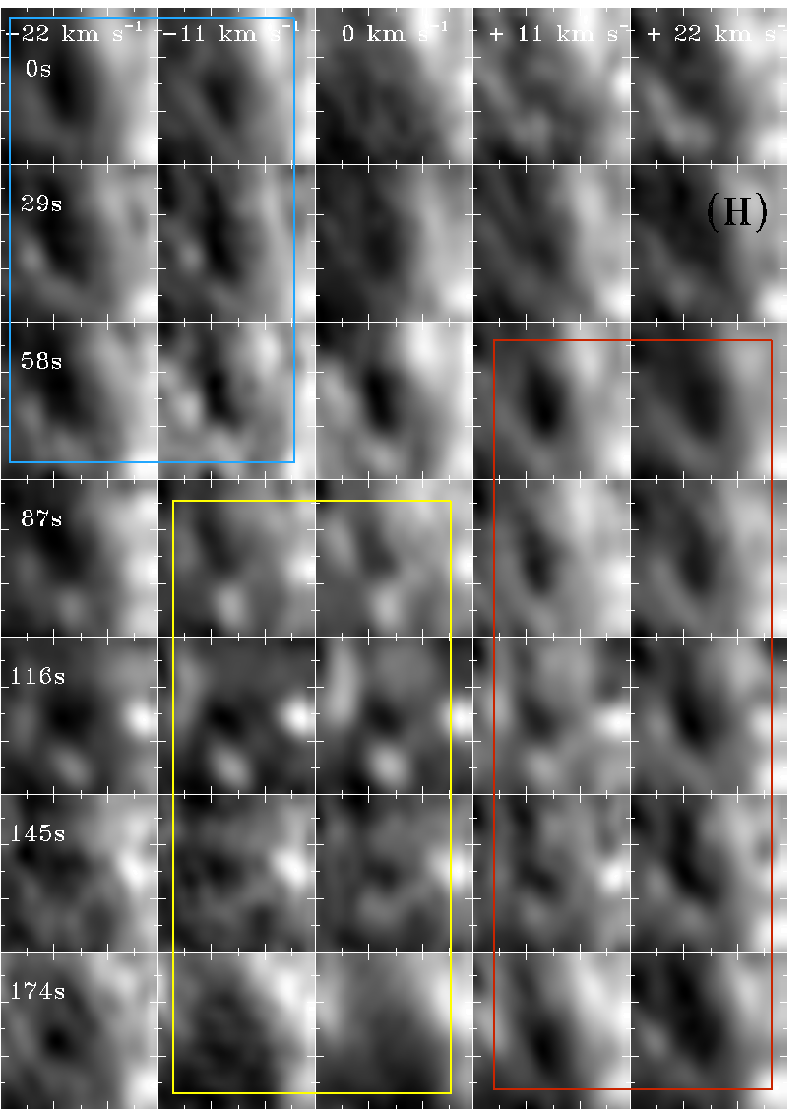} % H
\end{center}
\caption{Evolution of four different SSUB/SDF pairs (events E to H) in the same format as Fig.~\ref{trimedpanels}.}
\label{trimedpanels2}
\end{figure*}

\begin{figure}[!htb]
\centering
\includegraphics{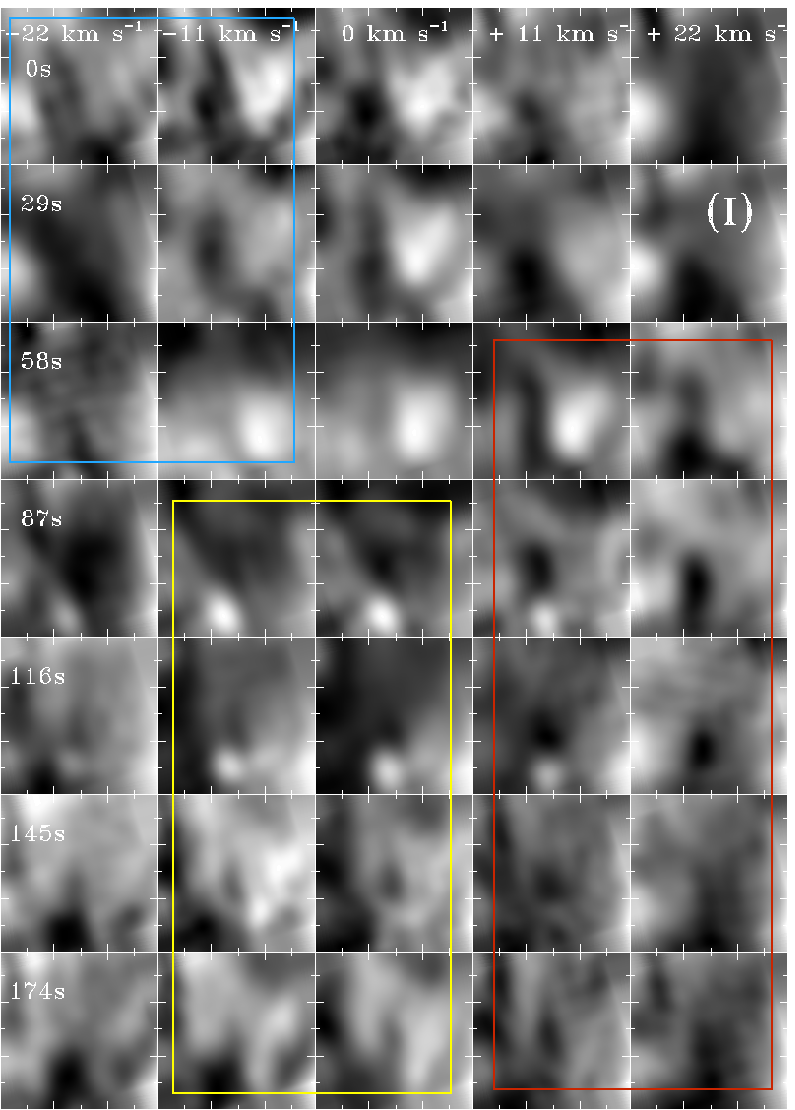} %I    % 
\caption{Evolution of SSUB/SDF pair (event I) in the same format as Fig.~\ref{trimedpanels}.}
\label{trimedpanels3}
\end{figure}

The spatial sampling was 0\farcs059 pixel$^{-1}$, with the spatial resolution not higher than the diffraction limit at this wavelength of 0\farcs18. The full FOV was 41$\times$41~Mm. Observations were taken at 15 wavelengths, from $-$290~m\AA\ to +290~m\AA\ in steps of 73~m\AA\ as measured from the averaged observed core of the \ion{Ca}{ii}~8542~\AA\ line in a quiet area at disk centre, and at $\pm$398~m\AA, $\pm$580~m\AA, and $\pm$942~m\AA. Full Stokes polarimetry was achieved by the two consecutive two-state LC modulators, producing four combinations of Stokes I, Q, U and V. These four combinations were then decomposed into I, Q, U and V using a demodulation matrix produced from a calibration of the optics after prime focus which, for these data, was taken less than 3 hours from the observations. The demodulation matrix used also includes a time-varying telescope model covering the optics before prime focus. The telescope model was produced from calibrations taken the same year and includes daily variations of the telescope polarization (see \citealt{2015A&A...573A..40D}, \citealt{2011A&A...534A..45S}, and \citealt{2010arXiv1010.4142S} for a complete description). 

Figure~\ref{context} introduces the nine SDF/SSUB events selected for detailed analysis. The spatial locations of these nine events are marked by the white boxes in the larger overview image (top-left panel), sampled in the far blue wing ($-$942~m\AA) of the \ion{Ca}{ii} $8542$ \AA\ line. The small panels (top-right) show the characteristic SDF--SSUB pairing at the moment of their clearest visibility. The SDF can be identified in each panel as the small dark blob and the associated SSUB is the brightening at its base. The bottom two panels of Fig.~\ref{context} display narrowband images of the sunspot taken at the line core (bottom-left) and at $+$145~m\AA\ (bottom-right) of the \ion{Ca}{ii} $8542$ \AA\ line at the times when two SSUBs (A \& H) were present. Thresholding is applied to saturate 30 \% of the pixels. Such thresholding is enough to make both the the SDFs and the SSUBs just visible to the eye even within the broad field-of-view.

\subsection{Inversions}

We used \verb|NICOLE| \citep{2015A&A...577A...7S},  which is a multi-purpose inversion code parallelised for both synthesis and inversions of Stokes profiles under non-LTE conditions. No custom modifications where introduced into the code pre compilation. The inversions were performed similarly to \cite{2017ApJ...845..102H} (from now on H2017) with some differences described in this section. The most significant difference is that, for this work, no spatial nor temporal binning was performed before inversions. We were able to invert the Stokes I profiles without such spatial and temporal binning with at least the same level of ``inversion noise'', i.e., without an increased level of very different atmospheres in adjacent pixels or failed convergences/fits. The resulting Stokes V fits were slightly worse when compared to H2017 and thus, together with the reduced signal to noise, we made the decision not to perform height dependent magnetic field fitting and abstain from interpreting the magnetic field results other than verifying that they are reasonable for sunspot data and in line with the average levels obtained in H2017.  We invert a 5\arcsec$\times$5\arcsec\ area around each SSUB  at the time of their detection and at 90 seconds before they occur. 

Nodes were placed regularly at equidistant heights. We used \verb|NICOLE's| native equation-of-state (EOS) as opposed to the more common choice of the Wittmann EOS (e.g. H2017). Each per spectral-scan run was performed using a common starting sunspot model, obtained by running NICOLE with the same parameters as those of cycle~1 in Table 1 with the quiet Sun ``FAL-C'' model \citep{1993ApJ...406..319F} as starting guess, and then averaged spatially using a boxcar average of 5 by 5 pixels and smoothing vertically by interpolating the atmosphere with a spline taking only every seventh point to produce a "cycle zero" starting model. The grid used and height ranges were the same as in \cite{2019A&A...627A..46B}. No averaging was performed in following cycles after producing this starting model. Two iterative cycles, where the cycle~1 atmospheres were used as starting inputs for cycle~2, were performed to fit the profiles. Such iterative procedure is an approach used in \verb|NICOLE| since \cite{2011A&A...529A..37S}, having been first proposed by \cite{1992ApJ...398..375R}, and the rationale can be found in those publications. 

The number of nodes per cycle are shown in Table~\ref{T1}. For cycle~2, even though nodes were placed for the magnetic field, the weights for Stokes I were five times higher than for Stokes V and ten times higher than for Stokes Q and Stokes U. 

This was to account for the difference in signal to noise across parameters and because the primary focus is on the flow structure of the atmosphere. The noise was initially estimated visually from the shape of the profiles where sharp wavelength to wavelength variations are likely noise. To estimate the signal to noise we used the root mean square of the difference between the profiles across umbra and penumbra and the same profiles smoothed with a boxcar three index wavelengths wide. These signal-to-noise estimate ratios computed to 0.85, 0.99, 2.76 and 193 for Q/I, U/I, V/I and I respectively. A more traditional quality measure is to take the standard deviation at continuum wavelengths in a quiet area where no signal is expected. This yielded $8.0\times 10^{-3}$, $5.7 \times 10^{-3}$, and $7.9\times 10^{-3}$ for Q/I, U/I, V/I respectively. However this measure is in the near absence of signal for Q, U and V and thus captures mostly noise from uncorrected cross-talk and CCD noise.

The wavelength reference for the core was set to the average of the quietest region in the FOV. This differs from the sometimes preferred choice of taking the umbra itself as a rest reference. As an indication that should be taken as a rough estimate due to the dynamic nature of the umbra, such calibration differs by 120~\ms{} from the average umbra at rest (i.e. calibration patch is 120~\ms{} blue-shifted when compared to the umbra), averaged over 14~scans spread over the time series and for a small square portion of the umbra, selected to just include the subfields H and F shown in Fig.~\ref{context} which are located in the darkest umbra region. The difference obtained in this way was surprisingly small. For reference, for photospheric lines, the convective blue shift of the quiet Sun can be 300~\ms{} higher than the umbra \citep{2018A&A...617A..19L}. In the chromosphere such comparisons are much more uncertain but we are confident that our calibration is appropriate for flows of the order of a few \kms{}.   
 
The following package options were selected when running \verb|NICOLE| . For the non-LTE loop, radiative transfer of five bound levels plus continuum of the \ion{Ca}{ii} ion in complete redistribution \citep{1984mrt..book..173S} was used. The starting LTE populations were determined using the \verb|MULTI| \citep{MatsUppsala1986} approach. The velocity-free approximation and a Gaussian quadrature with three angles were chosen for the rays in the non-LTE loop for cycle~1. For cycle~2, five angles were used for higher population accuracy. The cubic DELO-Bezier solver was selected for the radiative transfer \citep{2013ApJ...764...33D}. Isotopes for the Ca atom as in \cite{2014ApJ...784L..17L} were also included to guarantee good fits of the red wing. 

\begin{table}
\caption{Number of nodes for different parameters}             
\label{T1}      
\centering                          
\begin{tabular}{c c c}       
\hline\hline                 
Nodes cycle 1 & Nodes cycle 2 & Parameters \\ 
\hline                        % 
   5 & 8 &  Temperature \\
   3 & 5 & Velocity \\ 
   1 & 1 & Microturbulence \\
   0 & 1 & $B_{z}$ \\ 
   0 & 1 & $B_{x}$ \\ 
   0 & 1 & $B_{y}$ \\
   0 & 0 & Macroturbulence\\
\hline                                   
\end{tabular}
\end{table}

\section{Results}

\subsection{Spectral and spatial evolution}
\label{trimmed}

In Figs.~\ref{trimedpanels}, \ref{trimedpanels2} and \ref{trimedpanels3} we explore in more detail the progression of SDFs into SSUBs, first shown in \cite{2017A&A...605A..14N}. The intensity in the panels of the figures are scaled logarithmically due to the large differences in intensity caused by the umbral flashes and the SSUBs. The time and wavelength dependence of such large variations, their spatial propagation and the fact that the visibility of the SDFs themselves progresses from the blue to the red wings of the line with time makes it difficult to treat the space, time and spectral dimensions independently. However, when all dimensions are shown, with the spectral dimension along the $x$-axis and time progressing top-down along the $y$-axis, the full picture becomes evident as the behaviour is the same for all SDF/SSUB pairs. 
Each SDF/SSUB pair is shown in its own panel, across Figs.~\ref{trimedpanels}-\ref{trimedpanels3}, labelled from (A) to (I). For each pair, centring and interpolating rotation were applied so that each SDF is aligned along the $y$-axis, and so that each SSUB occurs close to the bottom of the FOV. This causes the $y$-direction to also be roughly the same as that of the closest penumbral filaments, to be encountered progressing higher along the $y$-axis (see Fig.\ref{context} for an idea of the distances to the penumbra). 

Each panel has a blue, a yellow and a red box highlighting three progressive stages of evolution common to every case. The blue box in the top-left marks the initial stage of the SDF. It simply selects the blue wing of the line and the first 58~s. Every SDF is first seen inside this box, almost always starting as a very small elongation, smaller than 0\farcs5, (e.g. panels D, F and H) or a circular shape also smaller than 0\farcs5 of radius (e.g. panels A and E), with the exception being panels B, G, and I where the SDF already displays a significant extension at start. At the bottom of the blue box, after 58~s, the extent of the SDF has increased to, or at least very close to, its full elongation, which varies from SDF to SDF, exceeding 1\farcs0 for SDFs A, D, and I. No SDF is visible for the first time-steps just right of the blue box, meaning that the Doppler shift of the SDF is towards the blue and that the SDF is moving towards the viewer (up from the solar surface as the sunspot is very close to disk centre) at the same time as the SDF extends up, compatible with an upward moving and extending column. In the final stages of evolution, delineated by the red box in the lower-right, we see the SDF contract from its maximum extent back to near no elongation in the red wing. At the start of this contraction for cases C, D, H and I, or a few time steps after for all other cases, one sees the self-descriptive SSUB at the very base of the SDF. The clearest signatures of the SSUBs occur in the inner blue wing delineated by the yellow box, but these events are often also visible as a faint brightenings in the +11~\kms{} bandpass (always with respect to the background in this visualisation).  

The final position of the SDF and of the SSUB is partially overlapping in all cases, with more of the SDF visible and less of the SSUB visible the further to the lower-right in the red box one looks. The panels in Figs.~\ref{trimedpanels}, \ref{trimedpanels2} and \ref{trimedpanels3} dramatically demonstrate this relationship.
For panels A, B, C, F, H and I, the presence of the broader flash is seen in the centre rows (between $t=58$~s and $t=116$~s) as an enveloping brightening in the background. That the flash engulfs the SDF in this way, before the SSUB happens, is important as the flash wave-front has an horizontal propagation component. That the UF has time to surround the SSUB location before it occurs indicates a phase difference between the two that is not due primarily to this horizontal propagation. %

As would be expected for features linked to SDFs (see, for example, \citealt{2013ApJ...776...56R}, \citealt{2015A&A...574A.131H}), some periodicity is detected in the SSUBs analysed here. Most notably, SSUBs A, C, and D occur at the same spatial location, 2.5 minutes and 6 minutes apart, respectively (counting from SSUB A), meaning the top two rows of the panel corresponding to SSUB D are the same (with a slight off-set) as the bottom two rows plotted for SSUB C. This 3-min periodicity is similar to that detected for umbral flashes. In addition to these co-spatial and obvious cases, an SDF is visible in the top right sub-field (i.e., the strongest red-shift and earliest time) for SSUB E, with no signature in the blue box in the top row, and a bright blue-shifted feature compatible with a SSUB is evident in the top row of the panel corresponding to SSUB B. It is thus likely that these SDF and SSUB signatures are part of the respectively preceding SDF/SSUB pairs, an indication of periodicity.

\subsection{Analysis of the $\lambda$-t diagrams}
\label{lambdat}

\begin{figure*}[!htb]
\begin{center} 
\includegraphics[]{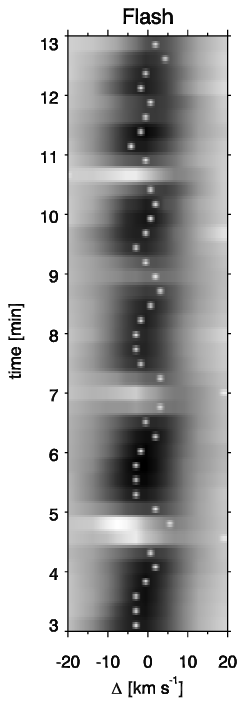} 
\includegraphics[]{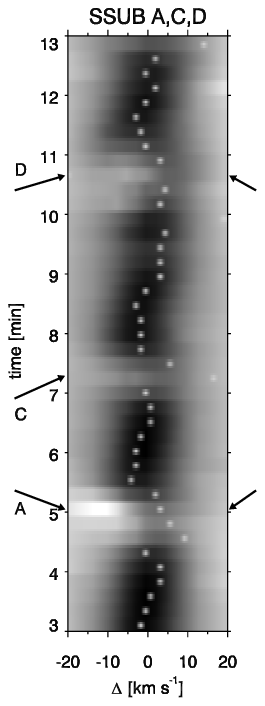} 
\includegraphics[]{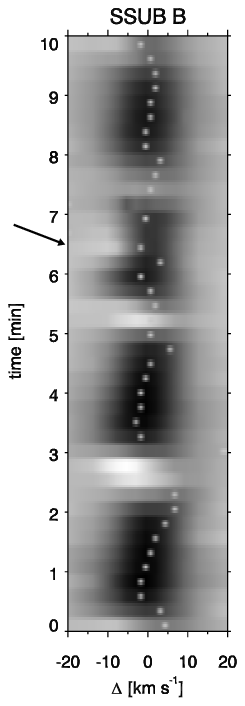}
\includegraphics[]{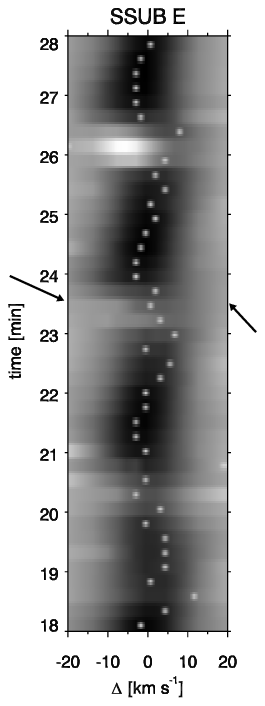} 
\end{center}
\caption{Spectral evolution in $\lambda-t$ diagrams showing the blue- to red-shift of the \CaIR{} line core between flashes (left panel) and SSUBs (right three panels). Cases from A to E as labelled. The black arrows from the left side indicate the moment of the SSUB and the right-side arrows indicate the darkening in the red wing caused by the collapsing SDF in the cases where this is clearly visible. The sampled wavelength range is clipped to $\pm$20~\kms. Linear interpolation to a regular grid of 36.5 m\AA\ (1.28~\kms{}) was used. The white dots indicate the Doppler-shift of the core of the line for each spectral scan as obtained by a three-point parabola fit using the core and the $\pm$~146~m\AA\ wavelengths. Note such quantity is unreliable in the presence of strong brightenings and is added here merely as an auxiliary to the eye when inspecting the profile evolution between bright events.} 
\label{lambdatplot1}
\end{figure*}

\begin{figure*}[!htb]
\begin{center} 
\includegraphics[]{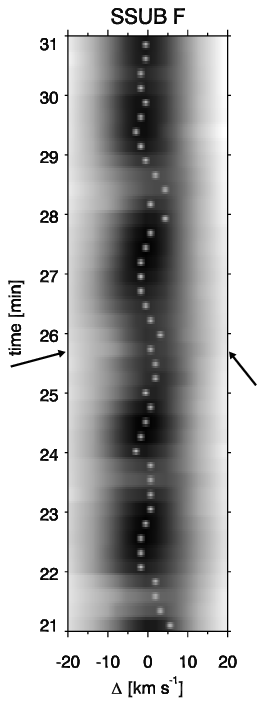}  
\includegraphics[]{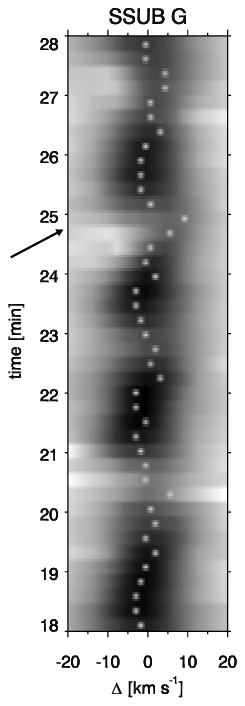}
\includegraphics[]{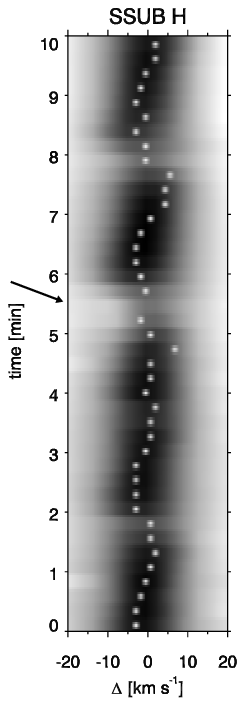}
\includegraphics[]{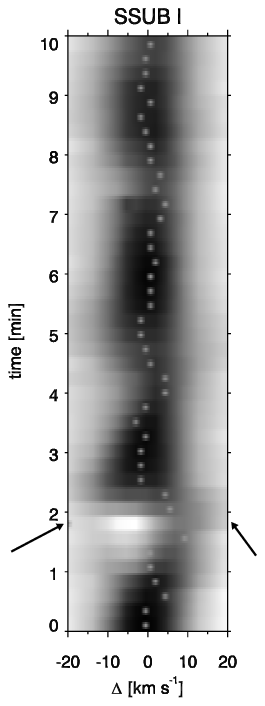}
\end{center}
\caption{Same as Fig.~\ref{lambdatplot1} but for cases from F to I.} 
\label{lambdatplot2}
\end{figure*}

In Figs.~\ref{lambdatplot1} and~\ref{lambdatplot2} we plot the  $\lambda$-t diagrams for all the SSUBs studied in this paper and an umbral flash in the vicinity of SSUB~A. Line profiles covering the wavelength range from -20 to +20~\kms{} are stacked in increasing time-steps. The white dots mark the Doppler-shift of the core of the line for each spectral scan as obtained by a three-point parabola fit using the core and the $\pm$~146~m\AA\ wavelengths. They serve primarily as a visual aids as they closely follows the behaviour of very visible dark inner wings of the line. The arrows on the left side of the diagrams show the location of the SSUBs whereas the right-side arrows show, whenever obvious, a dark excursion corresponding to the red-shifted SDF. For most SSUBs, such dark excursions into the red are visible at the same time as the SSUB. This matches the observation from the previous section analysis that SDFs and SSUBs overlap geometrically. The first $\lambda$-t diagram is that of an umbral flash neighbouring SSUB~A. The umbral flash is seen as a strong emission just before t=5~min. This same flash is visible in the the diagram for SSUBs~A, C, and D at the same t<5~min mark, but fainter. SSUB~A is seen as indicated by the respectively labelled arrow, slightly delayed from such flash front right after the t=5~min mark. This out of phase aspect is most striking for SSUB B, where both a preceding and a following flash front is visible, about one minute before and one minute after, respectively. SSUBs C, E, G, H and I all show a short delay from an immediately preceding flash front, most similar to that described for SSUB A, whereas for SSUB F a clear darkening is visible separating the preceding flash from the SSUB in time, more similar to the more out of phase SSUB~B. This out of phase behaviour with the broader flash pattern matches both the background brightening preceding the SSUBs, noted in the previous section, and the study based on light curves done in \cite{2017A&A...605A..14N}.

For umbral flashes, using the first column of Fig.~\ref{lambdatplot1} as an example, the quiescent dark line-core clearly progresses from a blue shift just after a brightening, to a clear red-shift just before the next umbral flash. This same behaviour occurs for SDFs, albeit harder to see in the $\lambda$-t diagrams due to the confounding signatures of the broader umbral flashes. In fact, the red-shifted absorption in cases A, D and E would seamlessly fit into such red-shifted progression if it were not for the immediately preceding strong flash front, which itself lacks a strong red-shifted absorption. For case I, the red excursion of the SDF seems to come after a more striking progression of the line core to the red. However, it is a strong sudden darkening much further into the red wing when compared to the preceding shift. In this case also, the red-shift of the core is indistinguishable from what would happen from a mere broadening of the blue emission feature. It is however, of remarkable progression and faint red-wing darkenings are seen long before SSUB itself around the t=1~min mark.   

The main and simplest pattern to be taken from these diagrams is that one sees the line-core and the inner line wings transitioning from a strong blue-shift just after an umbral flash, to a strong red-shift just before the next flash occurs, with almost the same behaviour in between SSUBs. This is not too dissimilar to the progression initially observed in the \ion{Ca}{ii}~K line \citep{1969SoPh....7..351B}, except in \CaIR{} this transition phase, the so called "quiescent phase", is simpler due to the absence of emission features.

\begin{figure}[!htb]
\begin{center} 

\includegraphics[width=4.4cm,clip=true,trim=0.3cm 0.1cm 0.2cm 0.3cm]{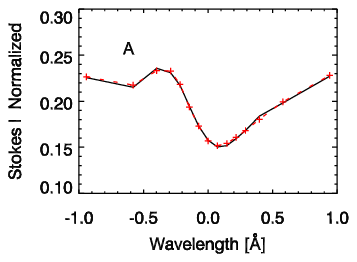} 
\includegraphics[width=4.4cm,clip=true,trim=0.3cm 0.1cm 0.2cm 0.3cm]{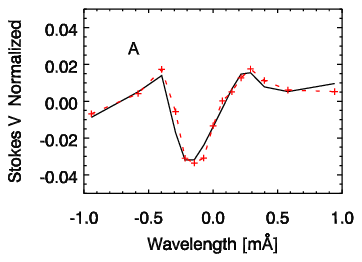} 
\includegraphics[width=4.4cm,clip=true,trim=0.3cm 0.1cm 0.2cm 0.3cm]{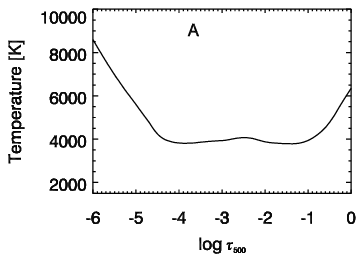}
\includegraphics[width=4.4cm,clip=true,trim=0.3cm 0.1cm 0.2cm 0.3cm]{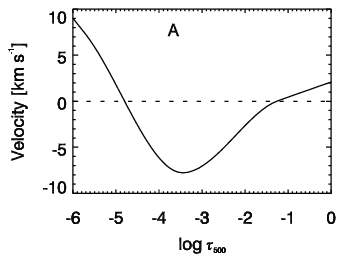} 

\includegraphics[width=4.4cm,clip=true,trim=0.3cm 0.1cm 0.2cm 0.3cm]{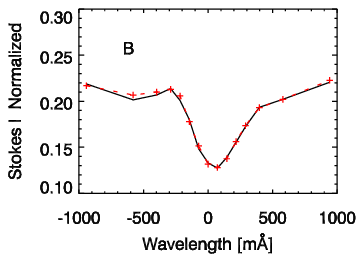}
\includegraphics[width=4.4cm,clip=true,trim=0.3cm 0.1cm 0.2cm 0.3cm]{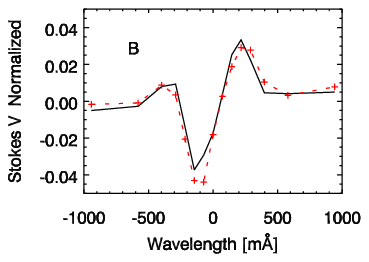} 
\includegraphics[width=4.4cm,clip=true,trim=0.3cm 0.1cm 0.2cm 0.3cm]{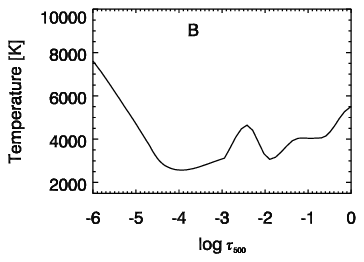}
\includegraphics[width=4.4cm,clip=true,trim=0.3cm 0.1cm 0.2cm 0.3cm]{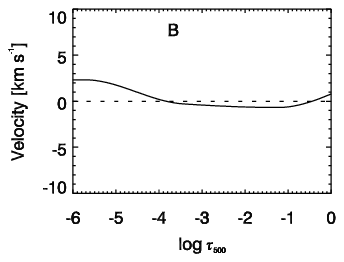}

\includegraphics[width=4.4cm,clip=true,trim=0.3cm 0.1cm 0.2cm 0.3cm]{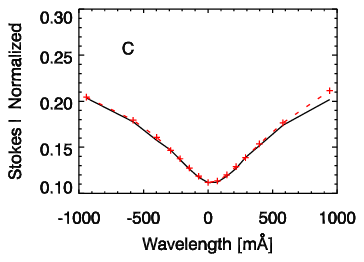}
\includegraphics[width=4.4cm,clip=true,trim=0.3cm 0.1cm 0.2cm 0.3cm]{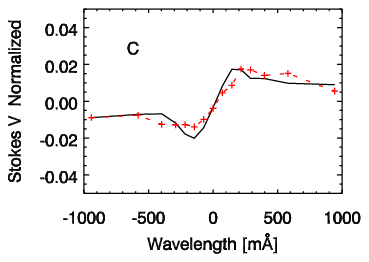}  
\includegraphics[width=4.4cm,clip=true,trim=0.3cm 0.1cm 0.2cm 0.3cm]{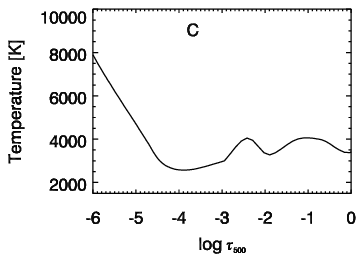}
\includegraphics[width=4.4cm,clip=true,trim=0.3cm 0.1cm 0.2cm 0.3cm]{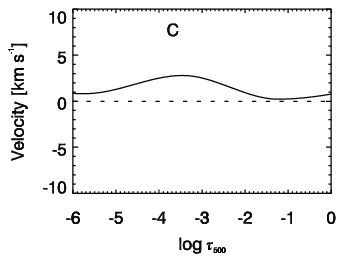}

\end{center}
\caption{Inversion results for three SSUBs (A, B, and C as labelled) and respective fits. Stokes I and V profiles from the inversion procedure are shown by a solid black line, while the observed profiles are plotted by a red dashed line connecting the specific observed wavelengths marked with crosses. Inverted temperature and velocity are shown as a function of optical depth at 500~nm. A dashed black line marks the zero velocity level with negative values standing for upflows. The inverted magnetic field strengths were 1.4~kG, 1.4~kG, and 1.3~kG Gauss for cases A, B and C respectively.}
\label{fits2}
\end{figure}

\begin{figure}[!htb]
\begin{center} 

\includegraphics[width=4.4cm,clip=true,trim=0.3cm 0.1cm 0.2cm 0.3cm]{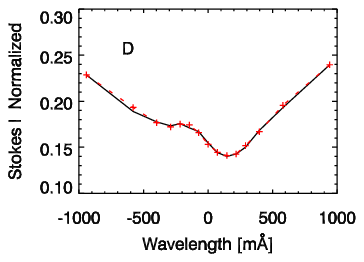}
\includegraphics[width=4.4cm,clip=true,trim=0.3cm 0.1cm 0.2cm 0.3cm]{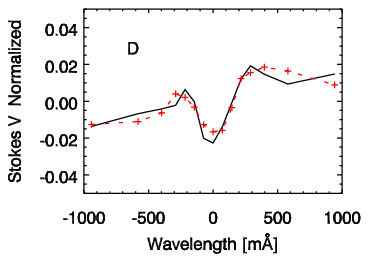} 
\includegraphics[width=4.4cm,clip=true,trim=0.3cm 0.1cm 0.2cm 0.3cm]{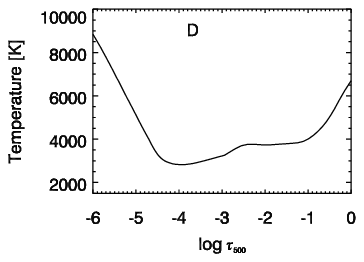}
\includegraphics[width=4.4cm,clip=true,trim=0.3cm 0.1cm 0.2cm 0.3cm]{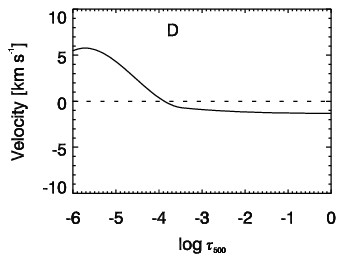}  

\includegraphics[width=4.4cm,clip=true,trim=0.3cm 0.1cm 0.2cm 0.3cm]{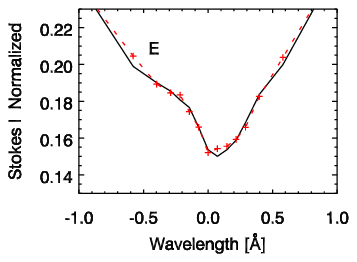}  
\includegraphics[width=4.4cm,clip=true,trim=0.3cm 0.1cm 0.2cm 0.3cm]{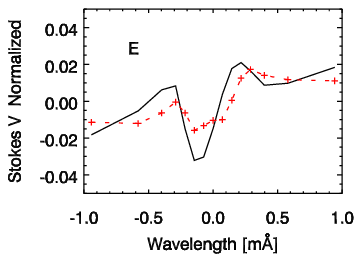} 
\includegraphics[width=4.4cm,clip=true,trim=0.3cm 0.1cm 0.2cm 0.3cm]{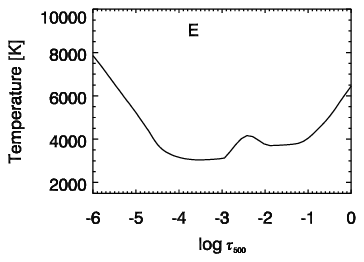}
\includegraphics[width=4.4cm,clip=true,trim=0.3cm 0.1cm 0.2cm 0.3cm]{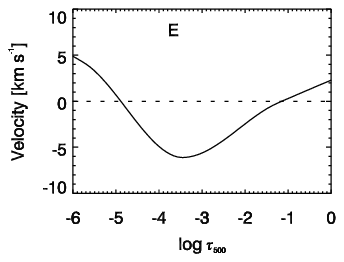}

\includegraphics[width=4.4cm,clip=true,trim=0.3cm 0.1cm 0.2cm 0.3cm]{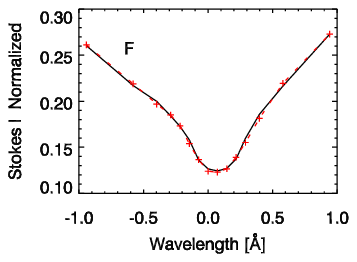}
\includegraphics[width=4.4cm,clip=true,trim=0.3cm 0.1cm 0.2cm 0.3cm]{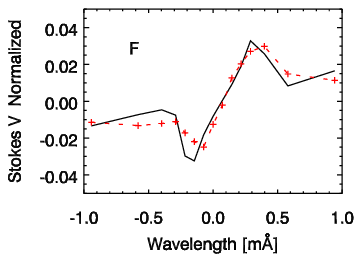} 
\includegraphics[width=4.4cm,clip=true,trim=0.3cm 0.1cm 0.2cm 0.3cm]{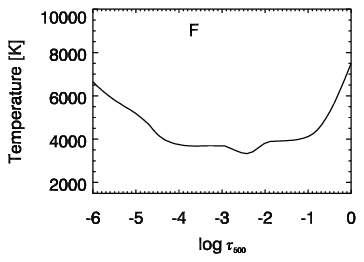}
\includegraphics[width=4.4cm,clip=true,trim=0.3cm 0.1cm 0.2cm 0.3cm]{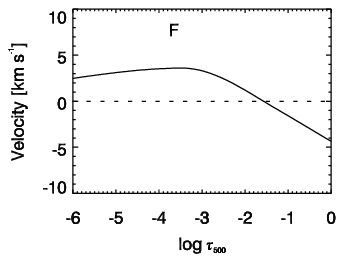}  

\end{center}
\caption{As Fig.~\ref{fits2} but for cases D, E and F. The inverted magnetic field strengths were 1.4~kG, 1.9~kG, and 1.2~kG Gauss, respectively.}
\label{fits3}
\end{figure}

\begin{figure}[!htb]
\begin{center} 

\includegraphics[width=4.4cm,clip=true,trim=0.3cm 0.1cm 0.2cm 0.3cm]{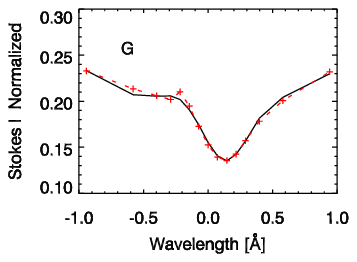}
\includegraphics[width=4.4cm,clip=true,trim=0.3cm 0.1cm 0.2cm 0.3cm]{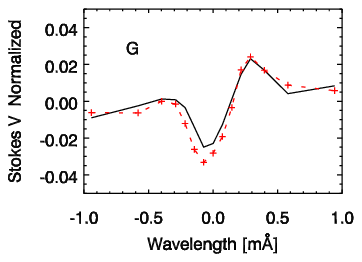} 
\includegraphics[width=4.4cm,clip=true,trim=0.3cm 0.1cm 0.2cm 0.3cm]{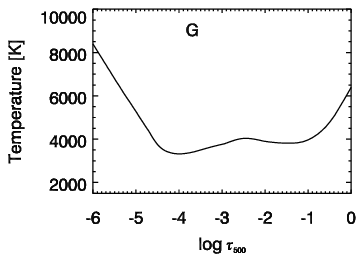} %
\includegraphics[width=4.4cm,clip=true,trim=0.3cm 0.1cm 0.2cm 0.3cm]{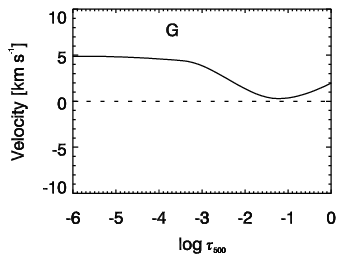}

\includegraphics[width=4.4cm,clip=true,trim=0.3cm 0.1cm 0.2cm 0.3cm]{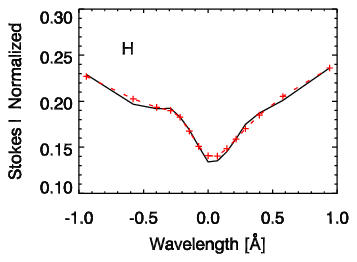} %
\includegraphics[width=4.4cm,clip=true,trim=0.3cm 0.1cm 0.2cm 0.3cm]{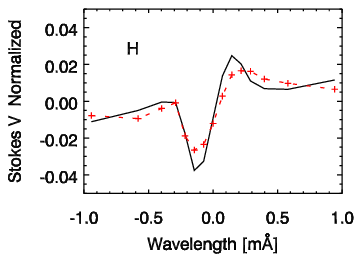} 
\includegraphics[width=4.4cm,clip=true,trim=0.3cm 0.1cm 0.2cm 0.3cm]{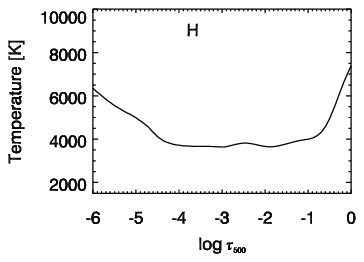}
\includegraphics[width=4.4cm,clip=true,trim=0.3cm 0.1cm 0.2cm 0.3cm]{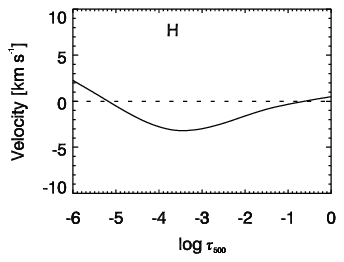}

\includegraphics[width=4.4cm,clip=true,trim=0.3cm 0.1cm 0.2cm 0.3cm]{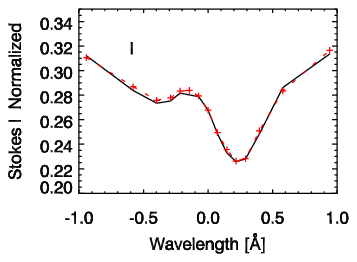} %1
\includegraphics[width=4.4cm,clip=true,trim=0.3cm 0.1cm 0.2cm 0.3cm]{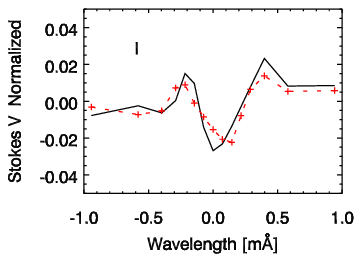} 
\includegraphics[width=4.4cm,clip=true,trim=0.3cm 0.1cm 0.2cm 0.3cm]{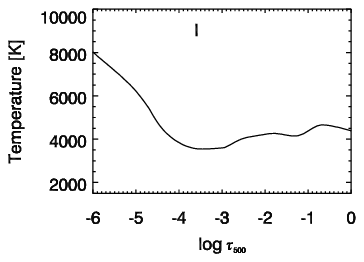}
\includegraphics[width=4.4cm,clip=true,trim=0.3cm 0.1cm 0.2cm 0.3cm]{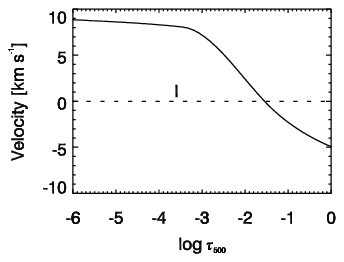}

\end{center}
\caption{As Fig.~\ref{fits2} but for cases G, H and I. The inverted magnetic field strengths were 1.3~kG, 1.4~kG and 1.3~kG Gauss, respectively.}
\label{fits4}
\end{figure}

\begin{figure}[!htb]
\begin{center} %
\includegraphics[width=4.4cm,clip=true,trim=0.3cm 0.1cm 0.3cm 0.3cm]{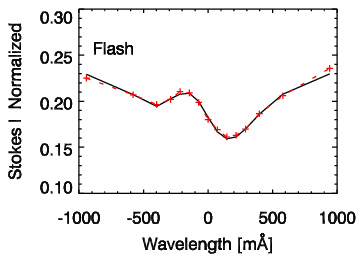}
\includegraphics[width=4.4cm,clip=true,trim=0.3cm 0.1cm 0.3cm 0.3cm]{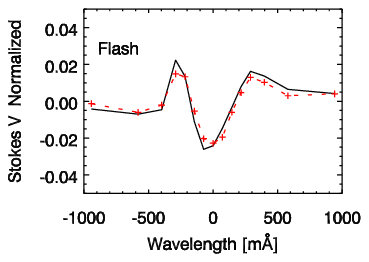} 
\includegraphics[width=4.4cm,clip=true,trim=0.3cm 0.1cm 0.3cm 0.3cm]{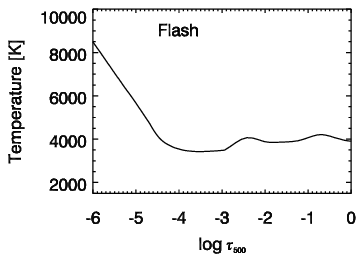}
\includegraphics[width=4.4cm,clip=true,trim=0.3cm 0.1cm 0.3cm 0.3cm]{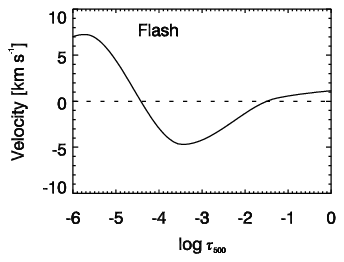} %

\includegraphics[width=4.4cm,clip=true,trim=0.3cm 0.1cm 0.3cm 0.3cm]{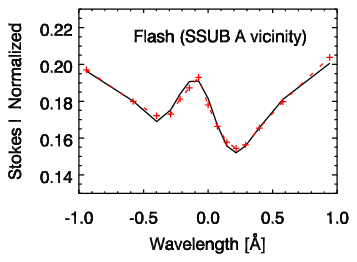}
\includegraphics[width=4.4cm,clip=true,trim=0.3cm 0.1cm 0.3cm 0.3cm]{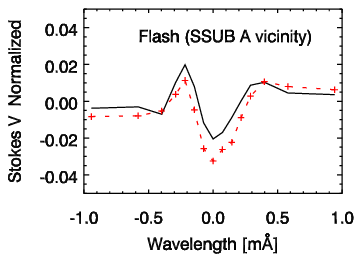} 
\includegraphics[width=4.4cm,clip=true,trim=0.3cm 0.1cm 0.3cm 0.3cm]{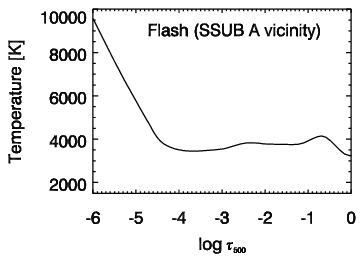}
\includegraphics[width=4.4cm,clip=true,trim=0.3cm 0.1cm 0.3cm 0.3cm]{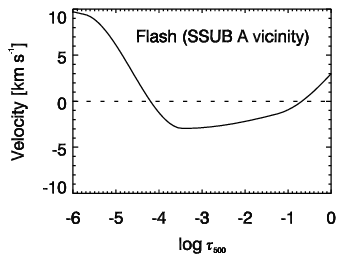} %

\end{center}
\caption{Profiles and inversion results for two umbral flashes showing both up and downflows along the vertical. Stokes I and V profiles, inverted temperature and velocity as a function of optical depth at 500~nm are plotted as labelled. Synthetic profiles from the inversion procedure are shown by a solid black line while the observed profiles are plotted in a dashed-red line connection symbols at the observed wavelengths. The inverted magnetic field (single node) for the top panel was 1.4~kG and 1.3~kG for the bottom panel. The bottom panel inverted flashed occurred in the vicinity of SSUB~A 30 seconds earlier.}
\label{flash_profiles}
\end{figure}

\subsection{Inversion Results}
\label{inversions}

Results from the inversions of the SSUBs are shown in Figs.~\ref{fits2}--\ref{fits4}, where the temperature and velocity stratification, as well as synthetic and observed spectral profiles are plotted for all SSUBs. The match between synthetic and observed profiles is consistently excellent for Stokes I and generally good for Stokes V, despite the respective level of noise encountered at these temporal and spatial scales. A red-shifted line-minimum and emission in the blue-wing is visible in the spectra of all SSUBs. This is similar to the profiles of umbral flashes (see Fig.~\ref{flash_profiles}) as was first pointed out by \cite{2017A&A...605A..14N}.

In the inverted SSUB atmospheres, the characteristic that is always present is that of downflows in the upper atmosphere, although it can be still be present as low in height as \logtau~=$-2$. One characteristic commonly present, but not always, is an enhancement in temperature just below this departure in velocities (between \logtau$=-2$ and $-3$). There may be some degeneracy between these two situations as both aspects contribute to an enhancement in the blue-wing of the line-profile. A relatively weaker temperature increase can be boosted by a stronger upper downflow (see later discussion in Section~\ref{opacitywindow}), but the strength of the flow should be constrained by the shape of the inner red wing and position of the line minimum (in addition to its impact on the blue-emission feature). 

The most common velocity stratification for SSUBs is that of a "counter-flow" structure as seen for cases A, B, D, E, F, H and I. These solutions have upflows at lower heights and downflows at higher heights. The exact depth of the steep gradient transition leading to a change of sign of the flow, as well as the amplitude of the flows, varies with case. The zero point where upflows will meet downflows can be as deep as \logtau=$-2$ (cases F or I) or as high as \logtau=$-5$ (cases A, E and H). Case D shows such a crossing at \logtau=$-4$ but the highest gradient in velocity is closer to \logtau=$-4.5$. Cases B and G do not show an upflow but both have atmospheres at rest in the lower layers. 

Inversions of neighbouring flash profiles also often show these "counter-flow" solutions, such as the two cases plotted in Fig.~\ref{flash_profiles}. Counter-flow solutions in umbral flashes tend to be quite similar to either of the two velocity stratifications in Fig.~\ref{flash_profiles}, with an upflow increasing in magnitude from the upper photosphere and peaking between \logtau$=-3$ and $-4$. Upflow magnitudes register more commonly around 2~\kms{} but can be as high as 5~\kms{}. It is important to note that such powerful upflows, topping at 5~\kms{} as in the top flash panel of Fig.~\ref{flash_profiles}, are the exception. For SSUBs, only SSUB A, the most powerful SSUB of the set, registers upflowing velocities exceeding 5~\kms{}, between \logtau$=-3$ and $-4$ with a top velocity of +8~\kms{}. The transition between up and downflows, for umbral flashes, occurs typically between \logtau$=-4$ and $-5$ as part of a steep gradient connecting the maximum upflow and the maximum downflow. If shocks are present, we would expect that this is the height at which they are occurring and that the steep gradient is capturing a smoothed version of the discontinuous shock velocity structure. Note that the steepest portion of the velocity stratification with height is not necessarily at the 0~\kms{} crossing point. % 

The inverted cubes do have columns showing either only upflows or only downflows, more in line with H2017 and the previous semi-empirical modelling for umbral flashes. The flash profiles selected for Fig.~\ref{flash_profiles} are also representative of the quality of the fits between the synthetic and observed profiles. Better fits are obtained, as are worse, but always in the form of a slight offset over the whole Stokes~V profile, as can be seen in the Stokes V profile of the bottom panel of Fig.~\ref{flash_profiles}, or a mismatch at the very peak of a sharp feature, as seen in the Stokes V blue peak of the top panel. Suggestively, this description and degree of match seems to us also applicable to the match of inverted Stokes V profiles and synthetics directly produced from simulations as tested by \cite{2019A&A...632A..75F}, for a slightly better sampling in wavelength selection around the core (55~m\AA\ versus 73~m\AA\ in this work). Some very bad fits contribute to the noise visible in the two dimensional maps analysed. In terms of hydrodynamical variables beyond temperature and velocity, as well as for response functions, our models are very similar to those of \cite{2019A&A...631L...5B} and thus we refer the reader to such work for their high-quality tables, both online and in print.

\begin{figure*}[!htb]
\begin{center}
  \begin{subfigure}[c]{0.03\textwidth}
    \textbf{A}
  \end{subfigure}\\
\includegraphics[height=3.0cm]{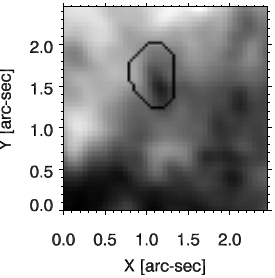} %
\includegraphics[height=3.0cm]{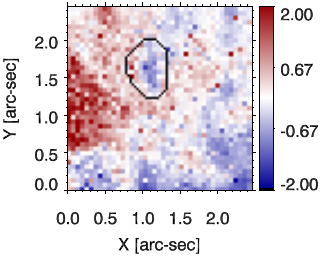} %
\hspace{5mm}
\includegraphics[height=3.0cm]{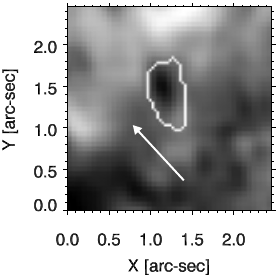}  % 
\includegraphics[height=3.0cm]{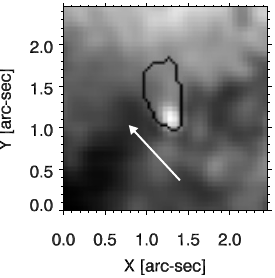}
\includegraphics[height=3.0cm]{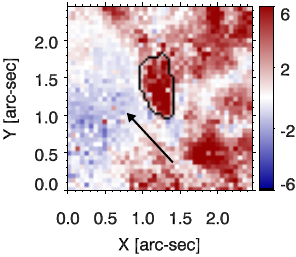}\\
  \begin{subfigure}[c]{0.03\textwidth}
    \textbf{B}
  \end{subfigure}\\
\includegraphics[height=3.0cm]{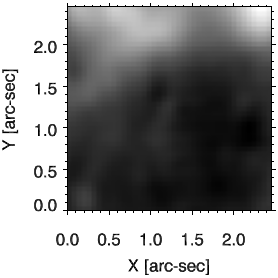} %
\includegraphics[height=3.0cm]{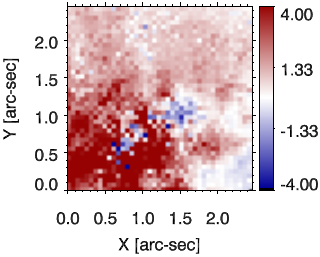}
\hspace{5mm}
\includegraphics[height=3.0cm]{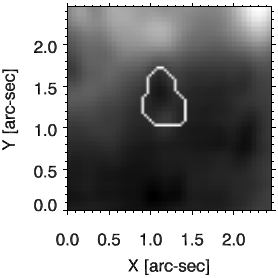}  %  B
\includegraphics[height=3.0cm]{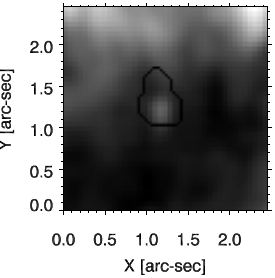}
\includegraphics[height=3.0cm]{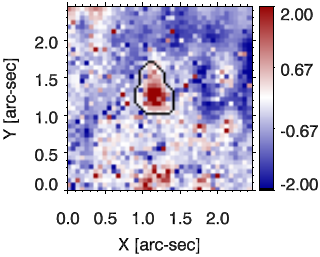} \\% 
  \begin{subfigure}[c]{0.03\textwidth}
    \textbf{C}
  \end{subfigure}\\
\includegraphics[height=3.0cm]{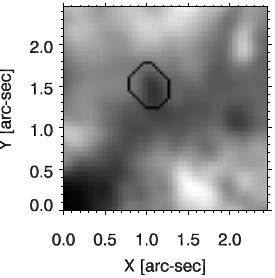} %0.5
\includegraphics[height=3.0cm]{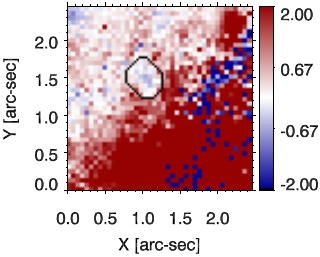}
\hspace{5mm}
\includegraphics[height=3.0cm]{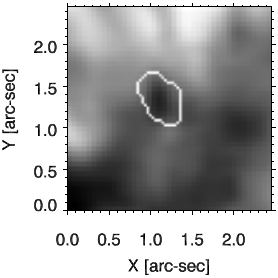}  %ssub3 00002.eps   C
\includegraphics[height=3.0cm]{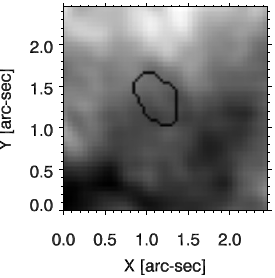} 
\includegraphics[height=3.0cm]{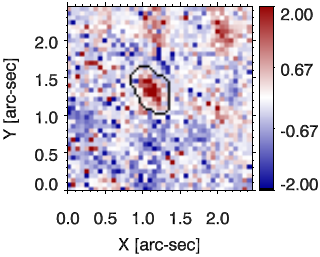}\\
  \begin{subfigure}[c]{0.03\textwidth}
    \textbf{D}
  \end{subfigure}\\
\includegraphics[height=3.0cm]{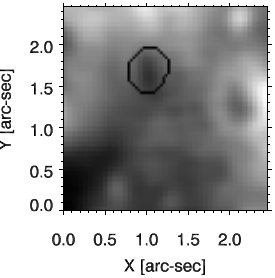}  % 
\includegraphics[height=3.0cm]{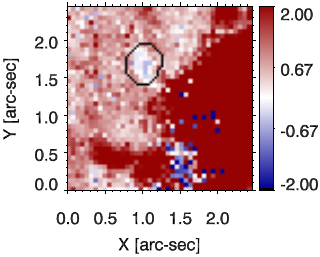}
\hspace{5mm}
\includegraphics[height=3.0cm]{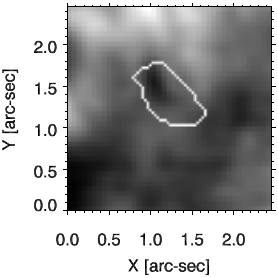}  % ssub4 00003.eps   D
\includegraphics[height=3.0cm]{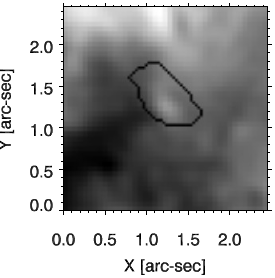}  
\includegraphics[height=3.0cm]{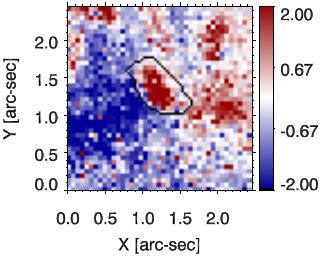}\\
\end{center}
\caption{Narrowband images and inverted velocity maps where downflowing structures matching the SDF/SSUB pairs are highlighted for four different cases (A to D as labelled). First column: upflowing stage of the SDFs as seen at $-$220~m\AA, 90~seconds before the SSUB. Second column: velocity map obtained from the respective inversions (i.e. from the full spectra and full FOV 90~s before the SSUBs) where the values were averaged between \logtau$=-4.5$ and $-5.5$. Third column: down phase of the SDF, consecutive with the SSUB, as seen at +220~m\AA . Fourth column: the SSUB, as visible at $-$220~m\AA. Last column: velocity map obtained from the inversions where the values were averaged between \logtau$=-4$ and $-5$. The contours highlight continuous regions automatically selected based on the presence of a flowing area that differs from the background. The arrow in case A highlights the "flash" location plotted in Figs.~\ref{lambdatplot1} and \ref{fits2}.} 
\label{2_d_Flows}
\end{figure*}

\begin{figure*}[!htb]
\begin{center}
\begin{subfigure}[!t]{0.03\textwidth}
    \textbf{E}
\end{subfigure}\\
\includegraphics[height=3.0cm]{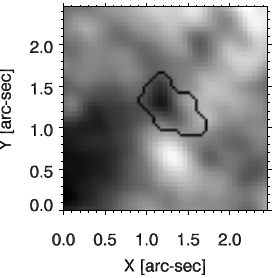}  % 
\includegraphics[height=3.0cm]{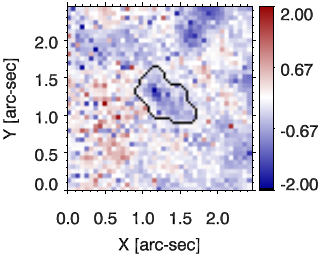}
\hspace{5mm}
\includegraphics[height=3.0cm]{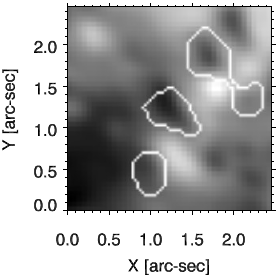} % ssub5 00004.eps    E
\includegraphics[height=3.0cm]{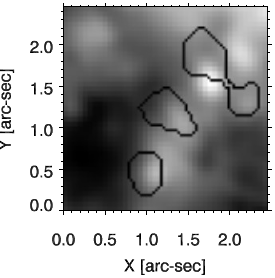}
\includegraphics[height=3.0cm]{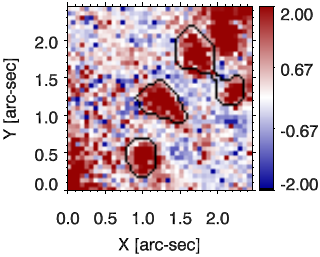}
\\

  \begin{subfigure}[c]{0.03\textwidth}
    \textbf{F}
  \end{subfigure}\\
\includegraphics[height=3.0cm]{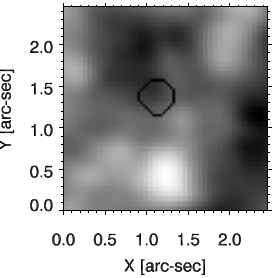}  % 
\includegraphics[height=3.0cm]{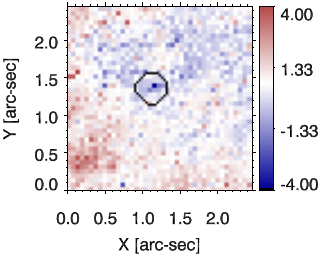}
\hspace{5mm}
\includegraphics[height=3.0cm]{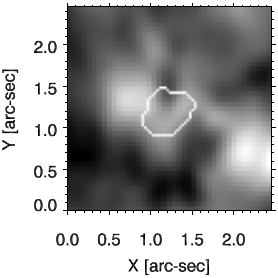} % ssub7 00006.eps    F
\includegraphics[height=3.0cm]{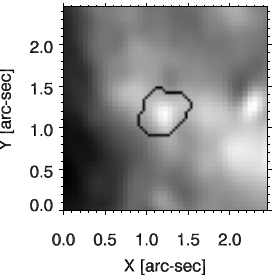}
\includegraphics[height=3.0cm]{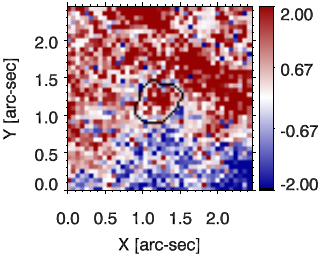} % 
\\

  \begin{subfigure}[c]{0.03\textwidth}
    \textbf{G}
  \end{subfigure}\\
\includegraphics[height=3.0cm]{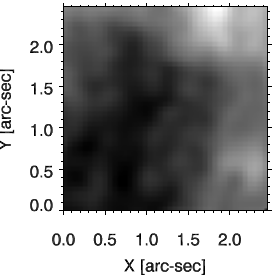}  
\includegraphics[height=3.0cm]{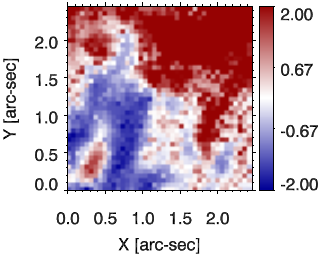} % 
\hspace{5mm}
\includegraphics[height=3.0cm]{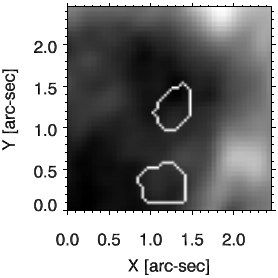} %
\includegraphics[height=3.0cm]{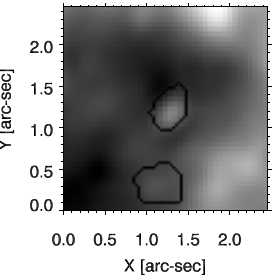}
\includegraphics[height=3.0cm]{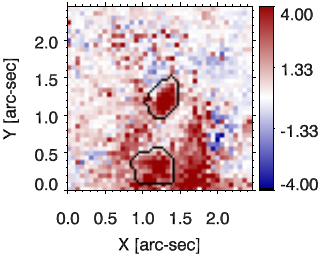}\\

  \begin{subfigure}[c]{0.03\textwidth}
    \textbf{H}
  \end{subfigure}\\
\includegraphics[height=3.0cm]{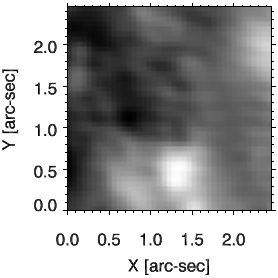}  
\includegraphics[height=3.0cm]{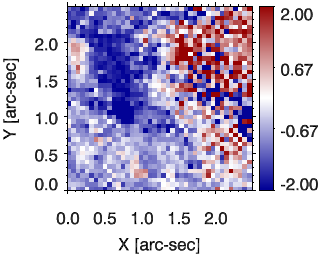} 
\hspace{5mm}
\includegraphics[height=3.0cm]{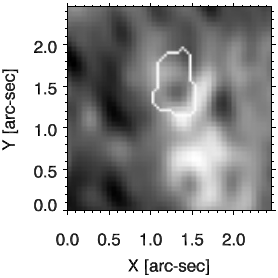}  
\includegraphics[height=3.0cm]{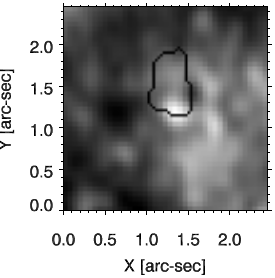}
\includegraphics[height=3.0cm]{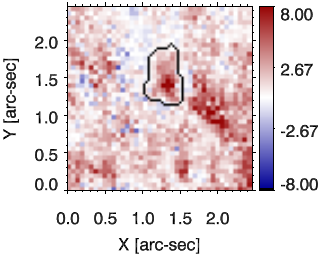}\\

\begin{subfigure}[c]{0.03\textwidth}
   \textbf{I}
\end{subfigure}\\
\includegraphics[height=3.0cm]{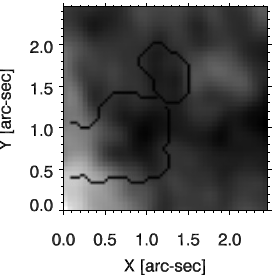}  
\includegraphics[height=3.0cm]{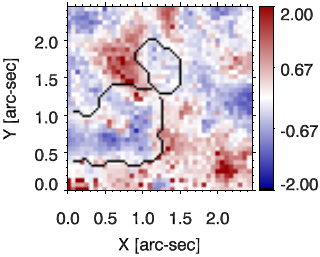} 
\hspace{5mm}
\includegraphics[height=3.0cm]{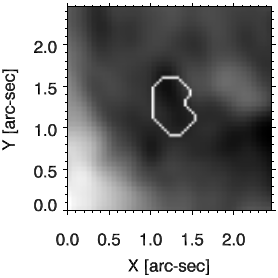} % 
\includegraphics[height=3.0cm]{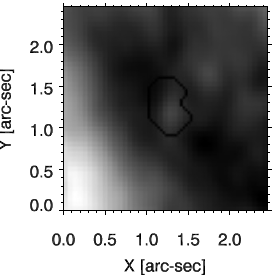}
\includegraphics[height=3.0cm]{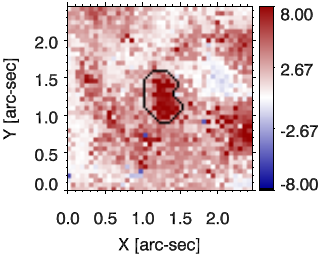}
\end{center}
\caption{Same as Fig.~\ref{2_d_Flows} but for SDF/SSUB pairs E through I.} 
\label{2_d_Flows_2}
\end{figure*}

\subsection{Analysis of the inverted velocity maps}
\label{2d}

In Figs.~\ref{2_d_Flows} and \ref{2_d_Flows_2} we show the average flow structure for \logtau$=-4$ to \logtau$=-5$ at the time of each SSUB detection, and the average flow structure 90~s earlier which captures the upflowing stages of the SDF. For the latter the averaging was done between \logtau$=-4.5$ to \logtau$=-5.5$. These optical depths were selected for being around the temperature minimum of most SSUBs and for contrast with the background for all features. The latter aspect was important for the visibility of the upflowing SDF stage in cases C and D, and for the downflowing SDF stage in case H. For some cases the magnitude of the inverted flows was stronger and re-scaling was necessary. This is reflected in the colour bar scale.

The black and white contours in Fig.~\ref{2_d_Flows} and \ref{2_d_Flows_2} are produced automatically using a combination of two thresholds in velocity. One threshold selects a contour in velocity that varies with case and thus with the strength of the flows (between 0.5 and 2 \kms{}), creating areas of similar flow structure. The second threshold selects such areas which contain the strongest flows visible in the FOV (one to three patches), downflows for SSUB stage and upflows for the upflowing SDF stage. A morphological open operation disconnects adjacent contours and a subsequent dilate extends the region by one pixel in each direction before the final contour is drawn. This procedure is intended to highlight regions of similar flows in the 2D maps.

Since each pixel is inverted independently there is some spatial variation in the obtained solutions due to noise, especially in the background umbra that surrounds the SDF/SSUB pairs. This, combined with the small feature size compared with that of the pixel scale, would cause any interpolating resampling of the post-inversion maps to alter the visible features significantly, generating artificial shapes and reducing contrast. For this reason, and unlike the analysis performed on the narrowband images of Section~\ref{trimmed}, no interpolation was performed for the velocity maps. A rotation to the closest 90 degrees was performed such that the SDF evolution is roughly the same as Figs.~\ref{trimedpanels}, \ref{trimedpanels2} and \ref{trimedpanels3} (i.e. extending up and falling down aligned with the y-axis). This y-axis direction roughly matches the orientation of the penumbral filaments. 

We find that, for all the cases, the SSUBs are downflowing throughout the whole body at the selected mapped heights. For each SSUB, the associated SDF is also consistently fitted with a downflow. More importantly, both features together form a continuous and coherent downflowing structure clearly visible in the two-dimensional Doppler maps. As captured by the contours, such structure traces the shape of the SSUBs and the shape of the SDF. Such continuous and coherent structure is always at contrast with the background and thus clearly traceable. The background has areas of downflows and upflows, often on opposite sides of the SSUB, with the upflowing areas tending to be areas post-flash and downflowing areas being pre-flash. The SSUB downflow is always stronger than its immediate (less than half an arcsec) surroundings. For cases B, C and D most of the background is either upflowing or at rest.

This spatial consistency between SSUBs and SDFs in the Doppler maps is remarkable as, again, each observed pixel is inverted independently with no spatial smoothing between inversion cycles. Adding to the remarkable spatial consistency of solutions, it is to be noted that SSUBs and SDFs are very different in appearance, evolution and spectrally, with SDFs essentially being a dark evolving fibril across the whole wavelength range and SSUBs bright specks centred in the inner blue wing of the line. As can also be seen from Figs.~\ref{2_d_Flows} and \ref{2_d_Flows_2}, the SSUBs, as in \cite{2017A&A...605A..14N} and as in Section~\ref{trimmed}, occur at the bottom of the body of the SDFs, during the downflowing phase of the SDF. 

In some of the panels one can see additional short-dynamic-fibril SSUB pairs that were not included in the main analysis (or not detected initially) that become evident when displaying data in this manner. This is the case with the non-centred contoured flows presented in panel E. The flow strength and the orientation of these extra features is similar to those of the main detections. This indicates that both the additional and the central features are close in phase and may share a causal relationship, with the obvious relation being the same wavefront that generates umbral flashes.   

Inspecting these maps we also find that the background areas that are upflowing generally had a flash in the past, and that downflowing or mixed flow areas are either about to flash or in the process of flashing. We selected one such upflowing background location, indicated with an arrow in Case A (in Fig.~\ref{2_d_Flows}) and plotted its $\lambda$-t diagram in Fig.~\ref{lambdatplot1} (labelled "flash"). As analysed in Sect.~\ref{lambdat}, we can clearly see the flash, just before the 5~min mark, just preceding the SSUB, followed by a clearly blue-shifted line-core. Based on such $\lambda$-t plots, we are confident that the inversions are capturing the blue-shifts post flash as upflows and we see no reason for why any other solution should be obtained for such a simple blue-shifted profile, likewise for downflowing areas pre-flash. 

\section{Discussion}

\subsection{The broader context of emission feature formation in the presence of strong flows}
\label{opacitywindow}

The results in this work add to a growing body of literature where Doppler-shifted emission features in chromospheric lines, that could be interpreted directly as the manifestation of emitting flows of similar shift, are best described by atmospheres where the strongest flow has the opposite direction of such Doppler-shift. Perhaps the first such case was the strongly downflowing model for the \ion{Ca}{ii}~H and K grains in the work of \cite{1970SoPh...11..347A} \citep[for a more recent discussion of this paper see][]{2010arXiv1012.1196R}, where it was argued that downflows of the order of 10-20 \kms\ in the upper line-forming layers (most contributing to the K$_3$ spectral feature) were a more likely description of reality than upflows close to the Doppler shift of the bright blue component (ranging from 3 to 7 \kms). This was due to the latter case leading to a remarkably low spatial variation when examining a wide FOV, whereas the downflowing models would allow more inhomogeneity, qualitatively matching the high variability in the position and width of the K$_3$ feature when compared to the position of the K$_2$ feature. Later \cite{1997ApJ...481..500C} solved the formation of such features with time-dependent models that included both upflows (1-2 \kms) in the mid-atmosphere, where the contribution for the blue-peak wavelength is highest, and strong downflows of the order of 10~\kms\ in the upper layers, where the K$_3$ contribution and opacity are highest. In such models the strong differential velocity would lead to an enhancement of the blue emission feature, which was present in the first place due to an increased coupling with the LTE source function during a density enhancement secondary to up-ward propagating waves, via increased collisions.

% we can now cite Souvik and Pasachoff 1969. 
Other works obtained similar emission features without the presence of a shock or density enhancement and merely velocity gradients. Such was the case discussed by \cite{1984mrt..book..173S}, who studied what would happen to the \ion{Ca}{ii}~H and K profiles in the case of a wave-train of upflows and downflows. In that work, K$_2$ emission features would alternate in strength, with the peak wavelength position being in opposite phase to the dominant flow in the atmosphere. The line-core minimum would shift in phase with the dominant flow and see an increase in intensity twice per wave period as the K$_3$ generating opacity was distributed in wavelength whereas the lower layers were not. The mechanism of this line-core enhancement was also described as a "reflector effect" in \cite{1981ApJ...249..720S}, where "unshifted" lower heights will illuminate the Doppler-shifted upper layers causing an enhancement in the line-core source function. The effect described in \cite{1984mrt..book..173S} was found to be contributing to the enhancement of a non-LTE emission feature in \cite{2015ApJ...810..145D}, displaying yet another example of a peak enhanced by velocity gradients connecting opposite flows at different heights. %LITES ALSO, but where? 

In other chromospheric features, \cite{2015ApJ...813..125K} have shown that such velocity gradients can shift the wavelength of maximum opacity to shorter or longer wavelengths generating red and blue line asymmetries, respectively. In such a scenario a red asymmetry is not necessarily associated with a plasma downflow and a blue asymmetry is not associated with a plasma upflow. \cite{2019A&A...631L...5B} found a remarkable direct relationship between the Doppler-shift of the line-minimum of \CaK{} in spicules and an enhancement of the emission feature in the opposite line wing. Thus the emission feature was suspected to be enhanced in a similar way to that put forward in \cite{1970SoPh...11..347A} and \cite{1997ApJ...481..500C}. Analysis based on k-mean clustering including the \Halpha\ profiles, analysis of the differential \Halpha\ profiles, and basic forward modelling led to the conclusion that the \CaK{} line-minimum was a good diagnostic of the real top-most spicule velocity with the emission feature enhancement being a secondary effect. Due to the evidence of visible background patterns when imaging the bright K$_2$ feature, and absent a convenient name for enhancement of spectral features upon removal of upper layer opacity, the term "opacity window" was coined.
 
For umbral flashes, and in the context of line-formation, that the enhanced inner-wing emission could be explained by a strong downflowing atmosphere, rather than an upflowing one, was first hypothesised by \cite{2003A&A...403..277R}, based on the similarity of the profiles with those of \cite{1970SoPh...11..347A}. The first inversions for umbral flashes that presented purely downflowing models were those of H2017, who first encountered strong downflowing semi-empirical solutions and argued that the downflows themselves might be associated with additional flow sources (additional to the steepening longitudinal oscillation at the heart of umbral flashes), such as deep-penetrating coronal rain, flocculent flows \citep{2012ApJ...750...22V}, inverse Evershed flows \citep[e.g.,][]{1975SoPh...43...91M,2019ApJ...874....6B}, or transition region downflows such as those observed by \cite{2015A&A...582A.116S}, \cite{2016A&A...587A..20C}, \cite{2018ApJ...859..158S}, and more recently by \cite{2020A&A...636A..35N}. The possibility of inverse Evershed flows contributing to portions of the umbra much further than the border with the penumbra is unlikely, but the impact of the other flow sources, at least when it comes to highly localised fine-structure, remains an emerging topic. 

Flows from a source other than the umbral oscillations fit into the interpretation within H2017 of an oscillating atmosphere, with up-ward moving shocks, but embedded in a bulk downflow (wherever a downflowing solution was present). Spatial asymmetries in such bulk flow could modulate the location of the shock front in terms of optical depth, leading to all the previously observed inhomogeneities \citep{2009ApJ...696.1683S,2013ApJ...776...56R,2013A&A...557A...5H,2014ApJ...787...58Y}. Following a similar investigation, \cite{2019A&A...627A..46B} found that such strong downflowing solutions, obtained from \CaIR{} inversions, also lead to good matches of synthetic \ion{Mg}{ii}~h and k profiles calculated from these inverted atmospheres when compared to IRIS observations. \cite{2020ApJ...892...49H} also obtains purely downflowing umbral flash solutions, and finds that most such solutions are preceded by an equally downflowing pre-flash atmosphere. Interestingly they also find upflowing umbral flash solutions, preceded by upflowing pre-flash atmospheres.
In this work we do not find a mix of purely upflowing or purely downflowing solutions but rather counter-flowing solutions whenever either flashes or SSUBs are present. This indicates that we are capturing the internal flows of the large-scale acoustic oscillations without the explanatory need of a baseline bulk flow. The consequences of this are discussed in the following sections. In the broader context of radiative transfer and complex line-profile generation this work adds another instance where velocity-gradient enhancements, via the opacity shifts they cause, are again important and lead to counter-intuitive features. Critical for the gradients obtained are the top-layer downflows.

\subsection{Consequence of strong colliding flows: density/temperature degeneracy}
\label{temperature}

From our multi-pronged analysis, showing clear evidence of a column of gas falling unto a brightening, and from the nature of shocks in general expected from previous literature and which always imply a local increase in density, we fully expect a stronger coupling of the source function to the local gas conditions to be occurring in reality for umbral flashes and SSUBs, secondary to a local increase in density and very similar to that present in \cite{1997ApJ...481..500C}. Unfortunately, \verb|NICOLE|, as all inversion codes presently in existence, cannot model such local enhancements in density due to the necessity of hydrostatic equilibrium in inversion mode. For this reason we do not trust the magnitude of the local enhancements in temperature. We further argue that present and past temperature enhancements obtained with semi-empirical inversion codes, for umbral flashes and shocks in general, should strictly be seen as what they are functionally to the profiles: enhancements in the source function at the heights at which they are present. If possible, accurate temperatures should be obtained via a simultaneous LTE diagnostic and such a diagnostic, if sampling the correct height, would allow for replacing the temperature as a free parameter with density. Then sufficiently constrained complete magnetohydrodynamic, NLTE semi-empirical modelling departing from hydrostatic equilibrium, i.e. the modelling of overdense clouds of gas at a known temperature, would be possible. The obvious observational tool for such diagnostic is the Atacama Large Milimeter/submillimeter Array (ALMA)  with its radio-frequency LTE emission diagnostics and solar observations capability \citep{2016SSRv..200....1W} at high resolution. \cite{2019A&A...627A..46B} argued, based on the temperature comparison between the ALMA observations of \cite{2019A&A...622A.150J} and the semi-empirical atmospheres of \cite{2019A&A...627A..46B}, that the formation of ALMA band 6 should be just at the top-most formation height of \CaIR{} during umbral flashes. 

\begin{figure*}[!htb]
\begin{center}
\includegraphics{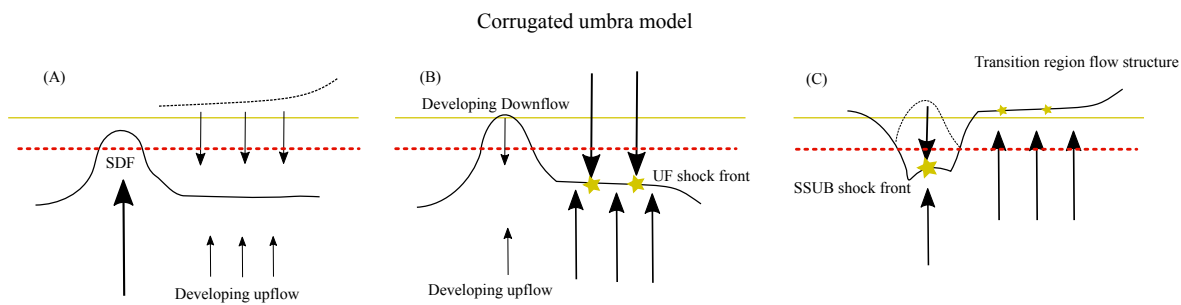}
\end{center}
\caption{Simple model of umbral flashes including SDFs and SSUBs as part of a broader corrugated structure. The different panels sequentially illustrate the cyclic state of the chromosphere of the umbra (i.e. time progresses from A to B to C to A and so on) with small arrows indicating a developing flow, large arrows a strong mass flow, the solid line indicating the surface where the velocity gradient is highest. The dashed black line in (A) and (C) indicates the previous position of the black line as reference. The yellow stars indicate shocks with size proportional to the strength of the shock, the yellow line indicates the absolute maximum height in the atmosphere that still contributes to line formation, and the dashed red line indicates the monochromatic $\tau=1$ surface for the line-minimum of \ion{Ca}{ii}~8542~\AA.}
\label{upss}
\end{figure*}

\subsection{Counter-flow solutions in SSUBs and umbral flashes}

The blue to red evolution of the line-core in between brightenings (discussed in Section.~\ref{lambdat}), the flow stratification of the SSUBs obtained via inversions  (discussed in Sect.~\ref{inversions}), and the two-dimensional patterns showing upflowing pre-flash areas as well as downflowing post-flash areas (discussed in Sect.~\ref{2d}), all together give us great confidence in the double-sign flow nature of the inverted solutions, for both SSUBs and umbral flashes. More specifically, both SSUBs and umbral flashes are occurring just as multiple signatures of an accelerating downflow come to an end, and just before signatures of strong upflows begin, with the inversions showing the two flows colliding at the moment of the brightening. We are thus confident that the origin of SSUBs involves downflowing material, from the return phase of the short-dynamic fibril, meeting new upflowing material. This is opposed to a reconnection scenario (as first proposed by \citealt{2013A&A...552L...1B}), although we do not exclude that other brightenings from reconnection might occur in the umbra for which there is recent evidence \citep{2020MNRAS.493.3036B}. Due to the range of results we are less confident on the exact amplitudes of the flows and their height locations, but these are necessarily affected by the oscillatory time-dependent nature of the phenomenon, and by the phase at which each SSUB was inverted. Refinement of amplitude and height locations can benefit from multi-diagnostic constraints for the top-most hydrodynamical boundary conditions (such as electron pressure) and improvements in velocity calibration other than just selecting a region deemed to be known in the FOV. However, we find these solutions to be unmatched in semi-empirical evidence strength due to their self-consistency when analysing the temporal, spatial and spectral evolution of the fine structure and broader flow patterns.   

\subsection{The empirical eyes}

Note that, for the first time, we capture the different flow stages of the umbra in between flashes using inversions. Earlier semi-empirical works put forward that most of the umbra could be slowly downflowing in a mass replenishing movement after a fast upward gas motions during the flash stage (\cite{2000Sci...288.1398S} and Model B of \cite{2001ApJ...550.1102S}), with the possibility that stronger downflows might aggregate in channels \citep{2006ApJ...640.1153C}. \cite{2001ApJ...550.1102S} also put forward an additional scenario where a purely downflowing and a purely upflowing component would be in close vicinity spatially, during a flash, but unresolved (Model A). The two-components in such earlier models were, however, not derived from the quiescent stage itself and not resolved in time. In later works, the spread of velocities of the quiescent phase, was small or non-existent (\citealt{2013A&A...556A.115D}, H2017, \citealt{2018A&A...619A..63J}) which, for those works where flashes were modelled strictly with upflows, meant that the mass balance was established out of sight or out of reach to the inversions, possibly in channels such as those proposed by \cite{2006ApJ...640.1153C}. The small velocity amplitudes of the quiescent phase obtained by recent inversion works may have been due to the mixing of profiles over the time it takes to perform a scan, a limitation that can affect the inverted velocities as identified by \cite{2018A&A...614A..73F}. Thus, it may be due to the improved scan cadence of this study, two times faster than in H2017, and faster than any other semi-empirical work to our knowledge, that we now can resolve the quiescent motions in spectro-polarimetric inversions. 

In H2017 it was speculated that downflowing atmospheres are successful in reproducing umbral flashes due to a radiative transfer effect (discussed in Sect.~\ref{opacitywindow}) that could be present in the inversions but not necessarily corresponding to reality, leading to a false degeneracy in solutions. That the solutions for the umbral flashes, in this work, show more commonly a counter-flow structure, makes us suspect that H2017 was capturing real downflows but just missing the lower upflowing component due to the aforementioned sampling effect. In fact, we suspect this might have been the case for the upflowing solutions therein also, i.e. \verb|NICOLE| would capture the upflowing section of the atmosphere but not the downflowing component, extrapolating the upflow structure into the upper layers of the model. We further suspect that this was the case for all the previous semi-empirical works using a scanning FPI, as they all captured either upflowing or downflowing solutions.

\subsection{The consequence for forward modelling work}

Magneto-acoustic waves generate flows at scales smaller than those of the wave's wavelength. While semi-empirical works have, up until now, not resolved such flows at all the stages of the oscillations behind umbral flashes, the full flow structure has been present in simulations since \cite{2010ApJ...722..888B}, with \cite{Felipe_2014} showing a very similar flow structure. In both such works, at the moment of the umbral flash, the waves cause the gas at height ranges exceeding 500~km to be either upflowing or downflowing. If \CaIR{} samples similar heights then it is not surprising that we are finally able to resolve such flows.

Indeed the disagreement between the counter-flowing solutions presented in this work and simulation studies comes down to whether or not the strong downflows in the upper layers of the shocked atmosphere are observable in the \CaIR{} line. In \cite{Felipe_2014} it was argued that for \CaIR{} only the upflowing section of the atmosphere would contribute to the line formation. However, if one disregards the height scale values, then the shape of the stratification and the amplitude values of the velocities obtained for the counter-flow solutions are remarkably similar to the umbral flash snapshots plotted in Fig.~16 of \cite{Felipe_2014}, for t=726 and t=744.

\subsection{Corrugated umbra model}

If flashes and SSUBs are so similar in nature, and if they are off by a phase corresponding to the time that the SDF takes to acquire descending momentum, then a scenario where the SDF/SSUB boundary is just a localised portion of the surface over which the umbral flash occurs becomes a strong and begging proposition. Such a surface would be highly corrugated and the greater geometric extent of the SDF could explain the delayed SSUB onset when compared to the rest of the flash generating surface, i.e., the compression or shock causing the flash to become visible would simply be delayed along an SDF. Since SSUBs occur at the bottom of the downflowing SDF, it is not a great leap to consider that the broader umbral flash is also likely occurring in the interface between the geometrically broader downflow and upflow pattern.
The source of the corrugation itself could be similar in nature to that in the MHD simulations of \cite{2011ApJ...743..142H}, performed in another context but of relevance for dynamic fibrils in general \citep{2006ApJ...647L..73H} as already noted by \cite{2013ApJ...776...56R} and \cite{2014ApJ...787...58Y}. 

A graphical depiction of this model is presented in Fig.~\ref{upss}. An isocontour of the umbra is depicted in three different instances with the solid black line indicating the strongest gradient region between upflows and downflows, the red line indicating the approximate formation height of the \CaIR{} line-core, and the yellow line indicating the maximum height contributing to line formation. 
At moment (A) no brightening is observed. The line-core in most of the umbra is shifted to the red except in the SDF where it is blue-shifted. In blue-wing narrowband imaging the SDF is visible as a fibril extending geometrically, if at a slight angle with the vertical depending on where in the umbra the SDF is located \citep{2013ApJ...776...56R}.  

At moment (B) the broader umbral flash occurs, in this case represented geometrically to the right of the SDF. This occurs as the upper downflows from moment A which are part of the restoring motion of the acoustic wave and strengthened by gravity, encounter the strengthening lower upflows generated by the same wave motion and leading to a region of high compression and/or a shock. The location of this compression is within the formation height of the \CaIR{} line, occurring close to the height of maximum gradient in velocity. The blue emission feature is formed by either a stronger coupling with the LTE source function (as in \citealt{1997ApJ...481..500C}) secondary to the increase in local density, due to an enhancement in temperature as captured in inverted solutions, or both. 

Note that while the compression region between up and downflows is likely a shock, it does not strictly have to be so. Whereas for \cite{2010ApJ...722..888B} the shock is within the line formation height, we see no reason why a scenario more like that of \cite{1997ApJ...481..500C} would not also generate umbral flash profiles in the real umbra and at the heights sampled by \CaIR{}, i.e. that a strong source function enhancement would happen already in a pre-shock compression feature as the acoustic wave is steepening and enhanced by high-up downflows, regardless of the presence of shocks higher up. In fact, our inverted models reproduce the observed profiles very well without a shock discontinuity, with the benefit of the temperature-density degeneracy previously discussed and with the convenient caveat that a shock may be present but unresolved at the heights for which we obtain the strongest velocity gradients.

In this model, the strong downflows contribute further to the enhancement of the source-function enhanced area where flows meet, as discussed in Sect.~\ref{opacitywindow}. Still at moment (B) the SDF has reached its maximum geometrical extent. A new upflow is developing in the deepest layers of the SDF as part of the 3-min oscillation. Straylight and horizontal illumination in the atmosphere from the neighbouring flash causes the flash to be faintly visible at the SDF location in $\lambda$-t diagrams but the SDF is clearly dark in two-dimensional narrowband imaging. 

At moment (C) the post-flash umbra is now upflowing and simple blue-shifted absorption line profiles are observed in \CaIR{}. Observations with IRIS, such as those of \cite{2014ApJ...786..137T} and \cite{2015ApJ...800..129M}, capture the remnants of the upwards-propagating wave in interface region diagnostics, above the limit detectable in \CaIR{} as depicted by the yellow line. However, the previously overextended SDF is now falling back down at full speed as it encounters the upflowing deep layers. This causes delayed and powerful compression, perhaps even a shock, which is then visible as a SSUB. The SSUB has umbral-flash like properties. The inversions of this feature capture the counter-flow structure with the top-most layers being downflowing. The $\lambda$-t plots show a brightening in the blue-wing, delayed or completely out of phase with the broader umbral flash with such delay also identifiable in two-dimensional maps showing the surrounding flash for preceding time frames, and a dark red-shifted signature from the collapsing SDF material. A slight dimple is depicted as the upflow gives rise to next upflowing SDF.

\section{Conclusions}

We confirm that SSUBs, first reported by \cite{2013A&A...552L...1B} as umbral microjets, and found by \cite{2017A&A...605A..14N} to have the same spectra and likely the same nature as umbral flashes, are not jets. As first put forward by \cite{2017A&A...605A..14N}, these are at least partially generated by downflows associated with the return flow phase of SDFs. In this paper we consistently fit both SDFs (primarily absorption features) and SSUBs (emission features) as a single structure that is continuous and in contrast to the remaining background. 

We propose that SSUBs are a modulation of the normal umbral flash front via a strong corrugation of the oscillating mass-flow fronts, which we can resolve in \CaIR{} due to the large scale of the acoustic oscillation at their origin. SDFs over extend geometrically in a corrugation like that of the two-dimensional \cite{2011ApJ...743..142H} models and, taking longer to fall back than the rest of the material, generate the SSUBs at either a slightly delayed time or completely out of phase with the broader umbral flash shock front. Behind this model is a coherent four-pronged collection of evidence in the form of spatio-temporal analysis of SDFs and SSUBs in context, the stratification of the inverted atmospheres per column, spatial analysis of the inverted Doppler patterns, and $\lambda$-t plots tracking the evolution of the spectral line. 

For the first time ever in a semi-empirical work we obtain atmospheric solutions, with strong support from the non-inversion analyses, that are both upflowing and downflowing along the same column during umbral flashes. Together with the tying in of umbral flashes with umbral chromospheric fine structure at the arcsec scale, manifested as SSUBs and SDFs, in a single model, we simultaneously understand how such fine structure occurs and do away with the need for return-flow channels, previously necessary to explain mass conservation \citep{2006ApJ...640.1153C}. This work shows that upflows and downflows occur at a depth such that both are essential for the line formation of chromospheric lines during umbral flashes, constraining future modelling work. This is shown  for the \ion{Ca}{ii}~8542 \AA\ line but necessarily also has to be the case for the \CaH{} and \CaK{} lines, due to their higher opacity. 

The space and time-dependent model put forward in this work, the corrugated umbra model, is the only model so far that simultaneously explains the properties of all dynamic chromospheric fine structure visible in the umbra and the umbral flashes themselves. It is a model with a semi-empirical origin and the main differences with regards to the simulation literature of acoustic generation of umbral flashes are the optical depth at which the umbral flashes are generated (deep enough that strong downflows are resolved) and the horizontal structure including corrugated features at an arcsec scale. This work ends up being mostly confirmatory of simulations but also constraining such future works as we expect future MHD simulations to include the spatial component and enough complexity that arcsec SDF/SSUBs pairs are generated. 

\begin{acknowledgements}
 We are deeply indebted to H. Socas-Navarro for continuously making his inversion code available to the community and to J.d.l.C.Rodr\'{i}guez for his contributions to such effort including the suit of accessory IDL routines. We appreciate the anonymous referee's comments in improving the manuscript.  This work would not be possible without the invaluable support of Peter S\"{u}tterlin at the SST. We would like to acknowledge the support by the Research Council of Norway, project number 250810, and through its Centers of Excellence scheme, project number 262622. This research also benefited from funding from the European Research Council (ERC) under the European Union’s Horizon 2020 research and innovation programme (grant agreement No. 682462). We have made use of the resources provided by the Vilje supercomputing facility at the NTNU, with the account NN2834K: Solar Atmospheric Modelling. C.J.N. and M.M. acknowledge support from STFC under grant No. ST/P000304/1. This research was supported by the SOLARNET project (www.solarnet-east.eu), funded by the European Commissions FP7 Capacities Program under the Grant Agreement 312495. The Swedish 1-m Solar Telescope is operated on the island of La Palma by the Institute for Solar Physics of Stockholm University in the Spanish Observatorio del Roque de los Muchachos of the Instituto de Astrof\'{i}sica de Canarias. The Institute for Solar Physics is supported by a grant for research infrastructures of national importance from the Swedish Research Council (registration number 2017-00625). This work made use of an IDL port of Robert Shine’s routines made by Tom Berger. These results were first presented at the IRIS-10 meeting in Bangalore, November 2019.
\end{acknowledgements}


\begin{thebibliography}{87}
%\expandafter\ifx\csname natexlab\endcsname\relax\def\natexlab#1{#1}\fi

\bibitem[{{Anan} {et~al.}(2019){Anan}, {Schad}, {Jaeggli}, \&
  {Tarr}}]{2019ApJ...882..161A}
{Anan}, T., {Schad}, T.~A., {Jaeggli}, S.~A., \& {Tarr}, L.~A. 2019, \apj, 882,
  161

\bibitem[{{Athay}(1970)}]{1970SoPh...11..347A}
{Athay}, R.~G. 1970, \solphys, 11, 347

\bibitem[{{Bard} \& {Carlsson}(2010)}]{2010ApJ...722..888B}
{Bard}, S. \& {Carlsson}, M. 2010, \apj, 722, 888

\bibitem[{{Beck} \& {Choudhary}(2019)}]{2019ApJ...874....6B}
{Beck}, C. \& {Choudhary}, D.~P. 2019, \apj, 874, 6

\bibitem[{{Beckers} \& {Tallant}(1969)}]{1969SoPh....7..351B}
{Beckers}, J.~M. \& {Tallant}, P.~E. 1969, \solphys, 7, 351

\bibitem[{{Bharti} {et~al.}(2013){Bharti}, {Hirzberger}, \&
  {Solanki}}]{2013A&A...552L...1B}
{Bharti}, L., {Hirzberger}, J., \& {Solanki}, S.~K. 2013, \aap, 552, L1

\bibitem[{{Bharti} {et~al.}(2020){Bharti}, {Sobha}, {Quintero Noda}, {Joshi},
  \& {Pandya}}]{2020MNRAS.493.3036B}
{Bharti}, L., {Sobha}, B., {Quintero Noda}, C., {Joshi}, C., \& {Pandya}, U.
  2020, \mnras, 493, 3036

\bibitem[{{Bloomfield} {et~al.}(2007){Bloomfield}, {Lagg}, \&
  {Solanki}}]{2007ApJ...671.1005B}
{Bloomfield}, D.~S., {Lagg}, A., \& {Solanki}, S.~K. 2007, \apj, 671, 1005

\bibitem[{{Bose} {et~al.}(2019{\natexlab{a}}){Bose}, {Henriques}, {Joshi}, \&
  {Rouppe van der Voort}}]{2019A&A...631L...5B}
{Bose}, S., {Henriques}, V. M.~J., {Joshi}, J., \& {Rouppe van der Voort}, L.
  2019{\natexlab{a}}, \aap, 631, L5

\bibitem[{{Bose} {et~al.}(2019{\natexlab{b}}){Bose}, {Henriques}, {Rouppe van
  der Voort}, \& {Pereira}}]{2019A&A...627A..46B}
{Bose}, S., {Henriques}, V.~M.~J., {Rouppe van der Voort}, L., \& {Pereira},
  T.~M.~D. 2019{\natexlab{b}}, \aap, 627, A46

\bibitem[{{Carlsson}(1986)}]{MatsUppsala1986}
{Carlsson}, M. 1986, Uppsala Astron.\ Obs.\ Report 33

\bibitem[{{Carlsson} \& {Stein}(1997)}]{1997ApJ...481..500C}
{Carlsson}, M. \& {Stein}, R.~F. 1997, \apj, 481, 500

\bibitem[{{Centeno} {et~al.}(2006){Centeno}, {Collados}, \& {Trujillo
  Bueno}}]{2006ApJ...640.1153C}
{Centeno}, R., {Collados}, M., \& {Trujillo Bueno}, J. 2006, \apj, 640, 1153

\bibitem[{{Centeno} {et~al.}(2005){Centeno}, {Socas-Navarro}, {Collados}, \&
  {Trujillo Bueno}}]{2005ApJ...635..670C}
{Centeno}, R., {Socas-Navarro}, H., {Collados}, M., \& {Trujillo Bueno}, J.
  2005, \apj, 635, 670

\bibitem[{{Chae} {et~al.}(2018){Chae}, {Cho}, {Song}, \&
  {Litvinenko}}]{2018ApJ...854..127C}
{Chae}, J., {Cho}, K., {Song}, D., \& {Litvinenko}, Y.~E. 2018, \apj, 854, 127

\bibitem[{{Chitta} {et~al.}(2016){Chitta}, {Peter}, \&
  {Young}}]{2016A&A...587A..20C}
{Chitta}, L.~P., {Peter}, H., \& {Young}, P.~R. 2016, \aap, 587, A20

\bibitem[{{Cho} {et~al.}(2019){Cho}, {Chae}, {Lim}, \&
  {Yang}}]{2019ApJ...879...67C}
{Cho}, K., {Chae}, J., {Lim}, E.-k., \& {Yang}, H. 2019, \apj, 879, 67

\bibitem[{{de la Cruz Rodr{\'{\i}}guez} {et~al.}(2015{\natexlab{a}}){de la Cruz
  Rodr{\'{\i}}guez}, {Hansteen}, {Bellot-Rubio}, \&
  {Ortiz}}]{2015ApJ...810..145D}
{de la Cruz Rodr{\'{\i}}guez}, J., {Hansteen}, V., {Bellot-Rubio}, L., \&
  {Ortiz}, A. 2015{\natexlab{a}}, \apj, 810, 145

\bibitem[{{de la Cruz Rodr{\'{\i}}guez} {et~al.}(2015{\natexlab{b}}){de la Cruz
  Rodr{\'{\i}}guez}, {L{\"o}fdahl}, {S{\"u}tterlin}, {Hillberg}, \& {Rouppe van
  der Voort}}]{2015A&A...573A..40D}
{de la Cruz Rodr{\'{\i}}guez}, J., {L{\"o}fdahl}, M.~G., {S{\"u}tterlin}, P.,
  {Hillberg}, T., \& {Rouppe van der Voort}, L. 2015{\natexlab{b}}, \aap, 573,
  A40

\bibitem[{{de la Cruz Rodr{\'{\i}}guez} \&
  {Piskunov}(2013)}]{2013ApJ...764...33D}
{de la Cruz Rodr{\'{\i}}guez}, J. \& {Piskunov}, N. 2013, \apj, 764, 33

\bibitem[{{de la Cruz Rodr{\'{\i}}guez} {et~al.}(2013){de la Cruz
  Rodr{\'{\i}}guez}, {Rouppe van der Voort}, {Socas-Navarro}, \& {van
  Noort}}]{2013A&A...556A.115D}
{de la Cruz Rodr{\'{\i}}guez}, J., {Rouppe van der Voort}, L., {Socas-Navarro},
  H., \& {van Noort}, M. 2013, \aap, 556, A115

\bibitem[{{De Pontieu} {et~al.}(2007){De Pontieu}, {Hansteen}, {Rouppe van der
  Voort}, {van Noort}, \& {Carlsson}}]{2007ApJ...655..624D}
{De Pontieu}, B., {Hansteen}, V.~H., {Rouppe van der Voort}, L., {van Noort},
  M., \& {Carlsson}, M. 2007, \apj, 655, 624

\bibitem[{{Felipe}(2019)}]{2019A&A...627A.169F}
{Felipe}, T. 2019, \aap, 627, A169

\bibitem[{{Felipe} \& {Esteban Pozuelo}(2019)}]{2019A&A...632A..75F}
{Felipe}, T. \& {Esteban Pozuelo}, S. 2019, \aap, 632, A75

\bibitem[{{Felipe} {et~al.}(2010){Felipe}, {Khomenko}, {Collados}, \&
  {Beck}}]{2010ApJ...722..131F}
{Felipe}, T., {Khomenko}, E., {Collados}, M., \& {Beck}, C. 2010, \apj, 722,
  131

\bibitem[{{Felipe} {et~al.}(2018{\natexlab{a}}){Felipe}, {Kuckein}, \&
  {Thaler}}]{2018A&A...617A..39F}
{Felipe}, T., {Kuckein}, C., \& {Thaler}, I. 2018{\natexlab{a}}, \aap, 617, A39

\bibitem[{{Felipe} {et~al.}(2014){Felipe}, {Socas-Navarro}, \&
  {Khomenko}}]{Felipe_2014}
{Felipe}, T., {Socas-Navarro}, H., \& {Khomenko}, E. 2014, \apj, 795, 9

\bibitem[{{Felipe} {et~al.}(2018{\natexlab{b}}){Felipe}, {Socas-Navarro}, \&
  {Przybylski}}]{2018A&A...614A..73F}
{Felipe}, T., {Socas-Navarro}, H., \& {Przybylski}, D. 2018{\natexlab{b}},
  \aap, 614, A73

\bibitem[{{Fontenla} {et~al.}(1993){Fontenla}, {Avrett}, \&
  {Loeser}}]{1993ApJ...406..319F}
{Fontenla}, J.~M., {Avrett}, E.~H., \& {Loeser}, R. 1993, \apj, 406, 319

\bibitem[{{Hansteen} {et~al.}(2006){Hansteen}, {De Pontieu}, {Rouppe van der
  Voort}, {van Noort}, \& {Carlsson}}]{2006ApJ...647L..73H}
{Hansteen}, V.~H., {De Pontieu}, B., {Rouppe van der Voort}, L., {van Noort},
  M., \& {Carlsson}, M. 2006, \apjl, 647, L73

\bibitem[{{Heggland} {et~al.}(2011){Heggland}, {Hansteen}, {De Pontieu}, \&
  {Carlsson}}]{2011ApJ...743..142H}
{Heggland}, L., {Hansteen}, V.~H., {De Pontieu}, B., \& {Carlsson}, M. 2011,
  \apj, 743, 142

\bibitem[{{Henriques}(2012)}]{2012A&A...548A.114H}
{Henriques}, V.~M.~J. 2012, \aap, 548, A114

\bibitem[{{Henriques} \& {Kiselman}(2013)}]{2013A&A...557A...5H}
{Henriques}, V.~M.~J. \& {Kiselman}, D. 2013, \aap, 557, A5

\bibitem[{{Henriques} {et~al.}(2017){Henriques}, {Mathioudakis},
  {Socas-Navarro}, \& {de la Cruz Rodr{\'{\i}}guez}}]{2017ApJ...845..102H}
{Henriques}, V.~M.~J., {Mathioudakis}, M., {Socas-Navarro}, H., \& {de la Cruz
  Rodr{\'{\i}}guez}, J. 2017, \apj, 845, 102

\bibitem[{{Henriques} {et~al.}(2015){Henriques}, {Scullion}, {Mathioudakis},
  {Kiselman}, {Gallagher}, \& {Keenan}}]{2015A&A...574A.131H}
{Henriques}, V.~M.~J., {Scullion}, E., {Mathioudakis}, M., {et~al.} 2015, \aap,
  574, A131

\bibitem[{{Houston} {et~al.}(2018){Houston}, {Jess}, {Asensio Ramos}, {Grant},
  {Beck}, {Norton}, \& {Krishna Prasad}}]{2018ApJ...860...28H}
{Houston}, S.~J., {Jess}, D.~B., {Asensio Ramos}, A., {et~al.} 2018, \apj, 860,
  28

\bibitem[{{Houston} {et~al.}(2020){Houston}, {Jess}, {Keppens}, {Stangalini},
  {Keys}, {Grant}, {Jafarzadeh}, {McFetridge}, {Murabito}, {Ermolli}, \&
  {Giorgi}}]{2020ApJ...892...49H}
{Houston}, S.~J., {Jess}, D.~B., {Keppens}, R., {et~al.} 2020, \apj, 892, 49

\bibitem[{{Jafarzadeh} {et~al.}(2019){Jafarzadeh}, {Wedemeyer}, {Szydlarski},
  {De Pontieu}, {Rezaei}, \& {Carlsson}}]{2019A&A...622A.150J}
{Jafarzadeh}, S., {Wedemeyer}, S., {Szydlarski}, M., {et~al.} 2019, \aap, 622,
  A150

\bibitem[{{Jess} {et~al.}(2015){Jess}, {Morton}, {Verth}, {Fedun}, {Grant}, \&
  {Giagkiozis}}]{2015SSRv..190..103J}
{Jess}, D.~B., {Morton}, R.~J., {Verth}, G., {et~al.} 2015, \ssr, 190, 103

\bibitem[{{Jess} {et~al.}(2019){Jess}, {Snow}, {Houston}, {Botha}, {Fleck},
  {Krishna Prasad}, {Asensio Ramos}, {Morton}, {Keys}, {Jafarzadeh},
  {Stangalini}, {Grant}, \& {Christian}}]{2020NatAs...4..220J}
{Jess}, D.~B., {Snow}, B., {Houston}, S.~J., {et~al.} 2019, Nature Astronomy,
  4, 220

\bibitem[{{Jess} {et~al.}(2017){Jess}, {Van Doorsselaere}, {Verth}, {Fedun},
  {Krishna Prasad}, {Erd{\'e}lyi}, {Keys}, {Grant}, {Uitenbroek}, \&
  {Christian}}]{2017ApJ...842...59J}
{Jess}, D.~B., {Van Doorsselaere}, T., {Verth}, G., {et~al.} 2017, \apj, 842,
  59

\bibitem[{{Joshi} \& {de la Cruz Rodr{\'\i}guez}(2018)}]{2018A&A...619A..63J}
{Joshi}, J. \& {de la Cruz Rodr{\'\i}guez}, J. 2018, \aap, 619, A63

\bibitem[{{Khomenko} \& {Collados}(2015)}]{2015LRSP...12....6K}
{Khomenko}, E. \& {Collados}, M. 2015, Living Reviews in Solar Physics, 12, 6

\bibitem[{{Krishna Prasad} {et~al.}(2015){Krishna Prasad}, {Jess}, \&
  {Khomenko}}]{2015ApJ...812L..15K}
{Krishna Prasad}, S., {Jess}, D.~B., \& {Khomenko}, E. 2015, \apjl, 812, L15

\bibitem[{{Kuridze} {et~al.}(2015){Kuridze}, {Mathioudakis}, {Sim{\~o}es},
  {Rouppe van der Voort}, {Carlsson}, {Jafarzadeh}, {Allred}, {Kowalski},
  {Kennedy}, {Fletcher}, {Graham}, \& {Keenan}}]{2015ApJ...813..125K}
{Kuridze}, D., {Mathioudakis}, M., {Sim{\~o}es}, P.~J.~A., {et~al.} 2015, \apj,
  813, 125

\bibitem[{{Leenaarts} {et~al.}(2014){Leenaarts}, {de la Cruz Rodr{\'{\i}}guez},
  {Kochukhov}, \& {Carlsson}}]{2014ApJ...784L..17L}
{Leenaarts}, J., {de la Cruz Rodr{\'{\i}}guez}, J., {Kochukhov}, O., \&
  {Carlsson}, M. 2014, \apjl, 784, L17

\bibitem[{{Lites}(1984)}]{1984ApJ...277..874L}
{Lites}, B.~W. 1984, \apj, 277, 874

\bibitem[{{Lites}(1992)}]{1992ASIC..375..261L}
{Lites}, B.~W. 1992, in NATO Advanced Science Institutes (ASI) Series C, Vol.
  375, NATO Advanced Science Institutes (ASI) Series C, ed. J.~H. {Thomas} \&
  N.~O. {Weiss}, 261--302

\bibitem[{{L{\"o}fdahl}(2002)}]{2002SPIE.4792..146L}
{L{\"o}fdahl}, M.~G. 2002, in Presented at the Society of Photo-Optical
  Instrumentation Engineers (SPIE) Conference, Vol. 4792, Society of
  Photo-Optical Instrumentation Engineers (SPIE) Conference Series, ed.
  {P.~J.~Bones, M.~A.~Fiddy, \& R.~P.~Millane}, 146--155

\bibitem[{{L{\"o}hner-B{\"o}ttcher} {et~al.}(2018){L{\"o}hner-B{\"o}ttcher},
  {Schmidt}, {Schlichenmaier}, {Doerr}, {Steinmetz}, \&
  {Holzwarth}}]{2018A&A...617A..19L}
{L{\"o}hner-B{\"o}ttcher}, J., {Schmidt}, W., {Schlichenmaier}, R., {et~al.}
  2018, \aap, 617, A19

\bibitem[{{Madsen} {et~al.}(2015){Madsen}, {Tian}, \&
  {DeLuca}}]{2015ApJ...800..129M}
{Madsen}, C.~A., {Tian}, H., \& {DeLuca}, E.~E. 2015, \apj, 800, 129

\bibitem[{{Maltby}(1975)}]{1975SoPh...43...91M}
{Maltby}, P. 1975, \solphys, 43, 91

\bibitem[{{Neckel}(1999)}]{1999SoPh..184..421N}
{Neckel}, H. 1999, \solphys, 184, 421

\bibitem[{{Nelson} {et~al.}(2017){Nelson}, {Henriques}, {Mathioudakis}, \&
  {Keenan}}]{2017A&A...605A..14N}
{Nelson}, C.~J., {Henriques}, V.~M.~J., {Mathioudakis}, M., \& {Keenan}, F.~P.
  2017, \aap, 605, A14

\bibitem[{{Nelson} {et~al.}(2020){Nelson}, {Krishna Prasad}, \&
  {Mathioudakis}}]{2020A&A...636A..35N}
{Nelson}, C.~J., {Krishna Prasad}, S., \& {Mathioudakis}, M. 2020, \aap, 636,
  A35

\bibitem[{{Pereira} \& {Uitenbroek}(2015)}]{Tiago2015RH}
{Pereira}, T. M.~D. \& {Uitenbroek}, H. 2015, \aap, 574, A3

\bibitem[{{Reid} {et~al.}(2017){Reid}, {Henriques}, {Mathioudakis}, {Doyle}, \&
  {Ray}}]{2017ApJ...845..100R}
{Reid}, A., {Henriques}, V., {Mathioudakis}, M., {Doyle}, J.~G., \& {Ray}, T.
  2017, \apj, 845, 100

\bibitem[{{Rimmele}(2008)}]{2008ApJ...672..684R}
{Rimmele}, T. 2008, \apj, 672, 684

\bibitem[{{Rouppe van der Voort} \& {de la Cruz
  Rodr{\'{\i}}guez}(2013)}]{2013ApJ...776...56R}
{Rouppe van der Voort}, L. \& {de la Cruz Rodr{\'{\i}}guez}, J. 2013, \apj,
  776, 56

\bibitem[{{Rouppe van der Voort} {et~al.}(2003){Rouppe van der Voort},
  {Rutten}, {S{\"u}tterlin}, {Sloover}, \& {Krijger}}]{2003A&A...403..277R}
{Rouppe van der Voort}, L.~H.~M., {Rutten}, R.~J., {S{\"u}tterlin}, P.,
  {Sloover}, P.~J., \& {Krijger}, J.~M. 2003, \aap, 403, 277

\bibitem[{{Ruiz Cobo} \& {del Toro Iniesta}(1992)}]{1992ApJ...398..375R}
{Ruiz Cobo}, B. \& {del Toro Iniesta}, J.~C. 1992, \apj, 398, 375

\bibitem[{{Rutten}(2010)}]{2010arXiv1012.1196R}
{Rutten}, R.~J. 2010, ArXiv e-prints [\eprint[arXiv]{1012.1196}]

\bibitem[{{Samanta} {et~al.}(2018){Samanta}, {Tian}, \& {Prasad
  Choudhary}}]{2018ApJ...859..158S}
{Samanta}, T., {Tian}, H., \& {Prasad Choudhary}, D. 2018, \apj, 859, 158

\bibitem[{{Scharmer}(1981)}]{1981ApJ...249..720S}
{Scharmer}, G.~B. 1981, \apj, 249, 720

\bibitem[{{Scharmer}(1984)}]{1984mrt..book..173S}
{Scharmer}, G.~B. 1984, {Accurate solutions to non-LTE problems using
  approximate lambda operators}, ed. W.~{Kalkofen}, 173--210

\bibitem[{{Scharmer}(2006)}]{2006A&A...447.1111S}
{Scharmer}, G.~B. 2006, \aap, 447, 1111

\bibitem[{{Scharmer} {et~al.}(2003{\natexlab{a}}){Scharmer}, {Bjelksjo},
  {Korhonen}, {Lindberg}, \& {Petterson}}]{2003SPIE.4853..341S}
{Scharmer}, G.~B., {Bjelksjo}, K., {Korhonen}, T.~K., {Lindberg}, B., \&
  {Petterson}, B. 2003{\natexlab{a}}, in Society of Photo-Optical
  Instrumentation Engineers (SPIE) Conference Series, Vol. 4853, Society of
  Photo-Optical Instrumentation Engineers (SPIE) Conference Series, ed.
  {S.~L.~Keil \& S.~V.~Avakyan}, 341--350

\bibitem[{{Scharmer} {et~al.}(2003{\natexlab{b}}){Scharmer}, {Dettori},
  {Lofdahl}, \& {Shand}}]{2003SPIE.4853..370S}
{Scharmer}, G.~B., {Dettori}, P.~M., {Lofdahl}, M.~G., \& {Shand}, M.
  2003{\natexlab{b}}, in Society of Photo-Optical Instrumentation Engineers
  (SPIE) Conference Series, Vol. 4853, Society of Photo-Optical Instrumentation
  Engineers (SPIE) Conference Series, ed. S.~L. {Keil} \& S.~V. {Avakyan},
  370--380

\bibitem[{{Scharmer} {et~al.}(2008){Scharmer}, {Narayan}, {Hillberg}, {de la
  Cruz Rodriguez}, {L{\"o}fdahl}, {Kiselman}, {S{\"u}tterlin}, {van Noort}, \&
  {Lagg}}]{2008ApJ...689L..69S}
{Scharmer}, G.~B., {Narayan}, G., {Hillberg}, T., {et~al.} 2008, \apjl, 689,
  L69

\bibitem[{{Schnerr} {et~al.}(2011){Schnerr}, {de La Cruz Rodr{\'\i}guez}, \&
  {van Noort}}]{2011A&A...534A..45S}
{Schnerr}, R.~S., {de La Cruz Rodr{\'\i}guez}, J., \& {van Noort}, M. 2011,
  \aap, 534, A45

\bibitem[{{Selbing}(2010)}]{2010arXiv1010.4142S}
{Selbing}, J. 2010, ArXiv e-prints, 1010.4142

\bibitem[{{Shine} {et~al.}(1994){Shine}, {Title}, {Tarbell}, {Smith}, {Frank},
  \& {Scharmer}}]{1994ApJ...430..413S}
{Shine}, R.~A., {Title}, A.~M., {Tarbell}, T.~D., {et~al.} 1994, \apj, 430, 413

\bibitem[{{Socas-Navarro}(2011)}]{2011A&A...529A..37S}
{Socas-Navarro}, H. 2011, \aap, 529, A37+

\bibitem[{{Socas-Navarro} {et~al.}(2015){Socas-Navarro}, {de la Cruz
  Rodr{\'{\i}}guez}, {Asensio Ramos}, {Trujillo Bueno}, \& {Ruiz
  Cobo}}]{2015A&A...577A...7S}
{Socas-Navarro}, H., {de la Cruz Rodr{\'{\i}}guez}, J., {Asensio Ramos}, A.,
  {Trujillo Bueno}, J., \& {Ruiz Cobo}, B. 2015, \aap, 577, A7

\bibitem[{{Socas-Navarro} {et~al.}(2009){Socas-Navarro}, {McIntosh}, {Centeno},
  {de Wijn}, \& {Lites}}]{2009ApJ...696.1683S}
{Socas-Navarro}, H., {McIntosh}, S.~W., {Centeno}, R., {de Wijn}, A.~G., \&
  {Lites}, B.~W. 2009, \apj, 696, 1683

\bibitem[{{Socas-Navarro} {et~al.}(2000{\natexlab{a}}){Socas-Navarro},
  {Trujillo Bueno}, \& {Ruiz Cobo}}]{2000ApJ...544.1141S}
{Socas-Navarro}, H., {Trujillo Bueno}, J., \& {Ruiz Cobo}, B.
  2000{\natexlab{a}}, \apj, 544, 1141

\bibitem[{{Socas-Navarro} {et~al.}(2000{\natexlab{b}}){Socas-Navarro},
  {Trujillo Bueno}, \& {Ruiz Cobo}}]{2000Sci...288.1398S}
{Socas-Navarro}, H., {Trujillo Bueno}, J., \& {Ruiz Cobo}, B.
  2000{\natexlab{b}}, Science, 288, 1398

\bibitem[{{Socas-Navarro} {et~al.}(2001){Socas-Navarro}, {Trujillo Bueno}, \&
  {Ruiz Cobo}}]{2001ApJ...550.1102S}
{Socas-Navarro}, H., {Trujillo Bueno}, J., \& {Ruiz Cobo}, B. 2001, \apj, 550,
  1102

\bibitem[{{Straus} {et~al.}(2015){Straus}, {Fleck}, \&
  {Andretta}}]{2015A&A...582A.116S}
{Straus}, T., {Fleck}, B., \& {Andretta}, V. 2015, \aap, 582, A116

\bibitem[{{Thomas}(1984)}]{1984A&A...135..188T}
{Thomas}, J.~H. 1984, \aap, 135, 188

\bibitem[{{Tian} {et~al.}(2014){Tian}, {DeLuca}, {Reeves}, {McKillop}, {De
  Pontieu}, {Mart{\'{\i}}nez-Sykora}, {Carlsson}, {Hansteen}, {Kleint},
  {Cheung}, {Golub}, {Saar}, {Testa}, {Weber}, {Lemen}, {Title}, {Boerner},
  {Hurlburt}, {Tarbell}, {Wuelser}, {Kankelborg}, {Jaeggli}, \&
  {McIntosh}}]{2014ApJ...786..137T}
{Tian}, H., {DeLuca}, E., {Reeves}, K.~K., {et~al.} 2014, \apj, 786, 137

\bibitem[{{Uitenbroek}(2001)}]{Uitenbroek2001}
{Uitenbroek}, H. 2001, \apj, 557, 389

\bibitem[{{van Noort} {et~al.}(2005){van Noort}, {Rouppe van der Voort}, \&
  {L{\"o}fdahl}}]{2005SoPh..228..191V}
{van Noort}, M., {Rouppe van der Voort}, L., \& {L{\"o}fdahl}, M.~G. 2005,
  \solphys, 228, 191

\bibitem[{{Vissers} \& {Rouppe van der Voort}(2012)}]{2012ApJ...750...22V}
{Vissers}, G. \& {Rouppe van der Voort}, L. 2012, \apj, 750, 22

\bibitem[{{Wedemeyer} {et~al.}(2016){Wedemeyer}, {Bastian}, {Braj{\v{s}}a},
  {Hudson}, {Fleishman}, {Loukitcheva}, {Fleck}, {Kontar}, {De Pontieu},
  {Yagoubov}, {Tiwari}, {Soler}, {Black}, {Antolin}, {Scullion}, {Gun{\'a}r},
  {Labrosse}, {Ludwig}, {Benz}, {White}, {Hauschildt}, {Doyle}, {Nakariakov},
  {Ayres}, {Heinzel}, {Karlicky}, {Van Doorsselaere}, {Gary}, {Alissandrakis},
  {Nindos}, {Solanki}, {Rouppe van der Voort}, {Shimojo}, {Kato},
  {Zaqarashvili}, {Perez}, {Selhorst}, \& {Barta}}]{2016SSRv..200....1W}
{Wedemeyer}, S., {Bastian}, T., {Braj{\v{s}}a}, R., {et~al.} 2016, \ssr, 200, 1

\bibitem[{{Wittmann}(1969)}]{1969SoPh....7..366W}
{Wittmann}, A. 1969, \solphys, 7, 366

\bibitem[{{Yurchyshyn} {et~al.}(2014){Yurchyshyn}, {Abramenko}, {Kosovichev},
  \& {Goode}}]{2014ApJ...787...58Y}
{Yurchyshyn}, V., {Abramenko}, V., {Kosovichev}, A., \& {Goode}, P. 2014, \apj,
  787, 58

\end{thebibliography}
\end{document}